\documentclass[aps,prx,reprint,a4paper,superscriptaddress,floatfix,amsmath,amssymb,amsfonts,noshowpacs,longbibliography]{revtex4-2} 
\usepackage{newtxtext,newtxmath} 
\usepackage[T1]{fontenc}
\usepackage[utf8]{inputenx} 
\usepackage{graphicx} 
\usepackage[dvipsnames]{xcolor} 
\usepackage{soul}  
\usepackage[textwidth=17.5cm,textheight=23.5cm,verbose,pdftex]{geometry} 
\usepackage[pdftex]{hyperref}

\hypersetup{pdfauthor={Uwe Jahn}, bookmarksnumbered=true, pdftitle={The carrier diffusion length in GaN---a cathodoluminescence study. I:~Temperature-dependent generation volume}, colorlinks, citecolor=blue, linkcolor=blue, urlcolor=blue}

\DeclareMathAlphabet{\mathcal}{OMS}{cmsy}{m}{n}

\begin{document} 
\title{Carrier diffusion in GaN---a cathodoluminescence study.\\ I:~Temperature-dependent generation volume} 
\author{Uwe Jahn} 
\author{Vladimir M. Kaganer} 
\affiliation{Paul-Drude-Institut für Festkörperelektronik, Leibniz-Institut im Forschungsverbund Berlin e.V., Hausvogteiplatz 5--7, 10117 Berlin, Germany} 
\author{Karl K. Sabelfeld} 
\author{Anastasya E. Kireeva} 
\affiliation{Institute of Computational Mathematics and Mathematical Geophysics, Russian Academy of Sciences, Lavrentiev Prosp.~6, 630090 Novosibirsk, Russia} 
\author{Jonas Lähnemann}
\email{laehnemann@pdi-berlin.de}
\author{Carsten Pfüller}
\author{Timur Flissikowski}
\author{Caroline Chèze} 
\author{Klaus Biermann} 
\author{Raffaella Calarco}
\altaffiliation{Present address: Istituto per la Microelettronica e Microsistemi, Consiglio Nazionale delle Ricerche, via del Fosso del Cavaliere 100, 00133~Roma, Italy} 
\author{Oliver Brandt} 
\email{brandt@pdi-berlin.de}
\affiliation{Paul-Drude-Institut für Festkörperelektronik, Leibniz-Institut im Forschungsverbund Berlin e.V., Hausvogteiplatz 5--7, 10117 Berlin, Germany}

\graphicspath{{./figs/}}

\begin{abstract} 
    The determination of the carrier diffusion length of semiconductors such as GaN and GaAs by cathodoluminescence imaging requires accurate knowledge about the spatial distribution of generated carriers. To obtain the lateral distribution of generated carriers for sample temperatures between 10 and 300~K, we utilize cathodoluminescence intensity profiles measured across single quantum wells embedded in thick GaN and GaAs layers. Thin (Al,Ga)N and (Al,Ga)As barriers, respectively, prevent carriers diffusing in the GaN and GaAs layers to reach the well, which would broaden the profiles. The experimental cathodoluminescence profiles are found to be systematically wider than the energy loss distributions calculated by means of the Monte Carlo program \texttt{CASINO}, with the width monotonically increasing with decreasing temperature. This effect is observed for both GaN and GaAs and becomes more pronounced for higher acceleration voltages. We discuss this phenomenon in terms of both, the electron-phonon interaction controlling the energy relaxation of hot carriers, and the shape of the initial carrier distribution. Finally, we present a phenomenological approach to simulate the carrier generation volume that can be used for the investigation of the temperature dependence of carrier diffusion.   
\end{abstract} 

\maketitle
\section{Introduction}

\label{sec:introduction} 
Scanning electron microscopy (SEM) enables a number of imaging, analytical and lithographical techniques with potentially high spatial resolution, as the electron beam can be focused down to 1~nm  in modern field-emission microscopes \cite{reimer_1998,goldstein_2003}. SEM is thus well suited and often the method of choice for the fabrication and investigation of nanostructures \cite{zhou_2007}. For a comprehensive physical and chemical characterization of these structures, various analytical techniques can be combined in a single instrument, such as secondary and backscattered electron imaging, electron backscatter diffraction, cathodoluminescence spectroscopy (CL) and energy dispersive x-ray spectroscopy (EDX) \cite{coenen_2017,lin_2017}. A spatial resolution corresponding to the minimum diameter of the focused electron beam can, however, only be achieved for secondary electron imaging. The actual spatial resolution of analytical techniques such as x-ray and CL spectroscopy is governed by the interaction of the primary high-energy electrons with matter. Elastic and inelastic scattering of these electrons leads to a cascade of subsequent excitations in the material such as excited atomic shell electrons (resulting in characteristic x-ray radiation), plasmons, and hot electron-hole pairs (resulting in CL emission after thermalization) within a generation volume that strongly depends on the energy of the impinging primary electrons \cite{goldstein_2003,edwards_2011}.

In the 1970s and 1980s, several empirical expressions have been proposed to approximate the spatial distribution of electron beam-generated excitations \cite{everhart_1971,fitting_1977,donolato_1981,oelgart_1984,werner_1988}. \citet{akamatsu_1989} were the first to derive an analytical expression of the generation volume based on Monte Carlo (MC) simulations. Since the 1990s, when personal computers became wide spread, user friendly MC programs for the simulation of the generation volume have been developed, which are now widely used in the SEM community \cite{holt_1994,hovington_1997,drouin_1997,drouin_2007,demers_2011}. Invariably, the generation volume is assumed to be given by the total electron energy loss distribution, regardless of the electron energies actually involved in the detected radiation. However, whereas characteristic x-ray radiation (detected in EDX) is only excited by electrons with a kinetic energy on the order of 1~keV and higher \cite{boyes_2000}, the radiative recombination of electron-hole pairs (detected in CL) only takes place between \emph{thermalized} electron and hole populations. It is intuitively clear that the generation volume should be larger in the latter case. At sufficiently low energies ($< 20$~eV), the thermalization of electrons and holes with the lattice occurs predominantly via the electron-phonon, and particularly the Fröhlich interaction \cite{llacer_1969,dapor_2012,dapor_2017}, which depends explicitly on temperature. MC programs such as \texttt{CASINO} \cite{hovington_1997,drouin_1997,drouin_2007,demers_2011}, however, do not allow us to consider this carrier thermalization process since they ignore temperature-dependent phenomena altogether.

In the present work, we are interested in the generation volume relevant for CL spectroscopy (thus including the thermalization of hot electrons and holes to the band edges) with the ultimate aim to reliably extract the carrier diffusion length in appropriately designed CL experiments. In the present article (the first of a series of three papers, hereafter referred to as CD1, CD2, and CD3), we first experimentally investigate the lateral extent of the CL generation volume in both GaN and GaAs. To this end, we use the CL emission from a single quantum well (QW) embedded in the GaN or GaAs matrix and clad by thin (Al,Ga)N or (Al,Ga)As barriers preventing carrier capture from the matrix by diffusion as originally proposed by \citet{bonard_1996}. In other words, the QW emission is produced exclusively by direct carrier generation within the QW structure, while thermalized carriers in the GaN or GaAs matrix cannot contribute to it. These results will be used in two subsequent studies of the temperature-dependent carrier diffusion in GaN presented in the companion papers CD2 \cite{brandt_2020} and CD3 \cite{lahnemann_2020}. 

For the present paper, we record CL line scans intersecting the single QWs embedded in GaN or GaAs for various temperatures. The resulting CL intensity profiles are compared with MC simulations utilizing \texttt{CASINO}~\cite{drouin_2007}. All experimental profiles turn out to be wider than the calculated ones, and their width increases further with decreasing temperature. Furthermore, the broadening at low temperatures becomes more pronounced for increasing acceleration voltages. The comparison of the results obtained on the two different materials helps to gain an understanding of the mechanisms giving rise to these effects. In particular, we attribute the broadening to the suppression of electron-phonon scattering processes with decreasing temperature, and its dependence on acceleration voltage to the evolution of the shape of the carrier distribution with the beam energy. Finally, we present a phenomenological model reproducing the experimental profiles with a single free parameter that we employ to determine the diffusion length in the following papers CD2 \cite{brandt_2020} and CD3 \cite{lahnemann_2020}.

\section{Preliminary considerations}

\label{sec:theory}

\begin{figure}
    \includegraphics[width=0.8\columnwidth] {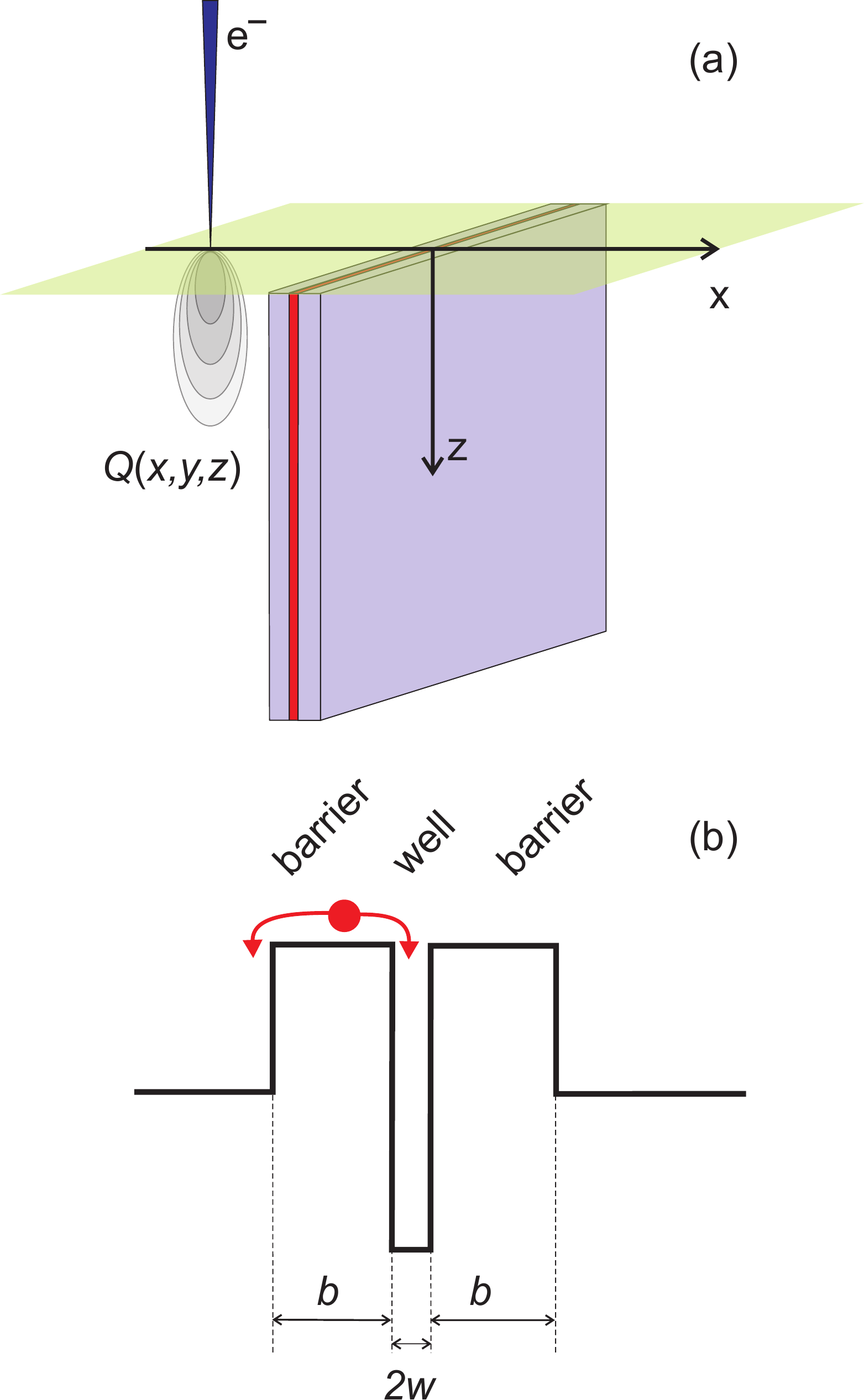}
   \caption{(a) Configuration of the CL experiment with the electron beam scanning across the QW of width $2w$ sandwiched between barriers of width $b$. (b) Sketch of the corresponding conduction band profile.}
   \label{fig1} 
\end{figure}

The configuration of our CL experiment is displayed schematically in Fig.~\ref{fig1}(a).  The focused electron beam impinges onto the cross-section of the sample along the \emph{z} direction. The beam and thus the carrier generation volume $Q(x,y,z)$ (source) is scanned across the QW along the $x$ axis normal to the well plane, with the center of the well being situated at $x=0$. The barriers of width $b$ [cf.\ Fig.~\ref{fig1}(b)] prevent a capture of thermalized diffusing carriers. In addition, the barrier height and width have to be chosen sufficiently large to ensure that carrier transport across the barrier by thermionic emission and tunneling can be safely neglected (for an estimate of the rate of bipolar carrier transport across the barrier, see Appendix \ref{Appendix:Thermionic}). In this case, only carriers excited within the barriers or directly within the QW will contribute to the detected CL signal of the QW.

According to the configuration of our experiment, we can consider the generation volume as a one-dimensional source after integrating over the $y$ and $z$ coordinates:

\begin{equation} 
    \bar{Q}(x)=\intop_{-\infty}^{\infty}dy\intop_{0}^{\infty}dz\,Q(x,y,z).
    \label{eq:1} 
\end{equation}
When this source is approaching the QW plane, the total flux of excited carriers captured by the well and thus the detected QW CL intensity $\mathcal{F}(x)$ can be written as a convolution

\begin{equation} 
\mathcal{F}(x)=\intop_{0}^{\infty}\bar{Q}(x-x')F(x')\,dx',
\label{eq:7} 
\end{equation} 
with the function $F(x')$ representing the fraction of carriers generated at $x'$ and reaching the well. $F(x')$ automatically includes all carriers excited directly in the QW. For carriers excited within the barriers, we consider three different cases:

(i) Carriers generated in the barriers are localized and do not reach the well. In this case $F(x')=F_{0}(x')=\delta(x')$ and the flux to the well is simply $\mathcal{F}(x)=\mathcal{F}_{0}(x)=\bar{Q}(x)$, where we have neglected the well width compared to the barrier width.

(ii) Carriers generated in the barriers can escape towards both the well and the matrix material. If the carrier diffusion length in the barrier is large compared to the barrier width, the fluxes to the well and to the matrix are weighted by the distance from the excitation position to the corresponding edges and can be written as:

\begin{equation} 
F(x)=F_{w}(x)=\begin{cases} 1-|x|/b, & |x|\leq b\\ 0, & |x|>b \end{cases}.
\label{eq:s4}
\end{equation}

(iii) All carriers generated in the barriers are captured by the well and produce the maximum possible flux as well as the maximum possible broadening of the energy loss profile. Then, the flux to the well is written as:

\begin{equation} 
F(x)=F_{m}(x)=\begin{cases} 1, & |x|\leq b\\ 0, & |x|>b \end{cases}.\label{eq:s5} 
\end{equation}

\begin{figure} 
    \includegraphics[width=0.85\columnwidth]{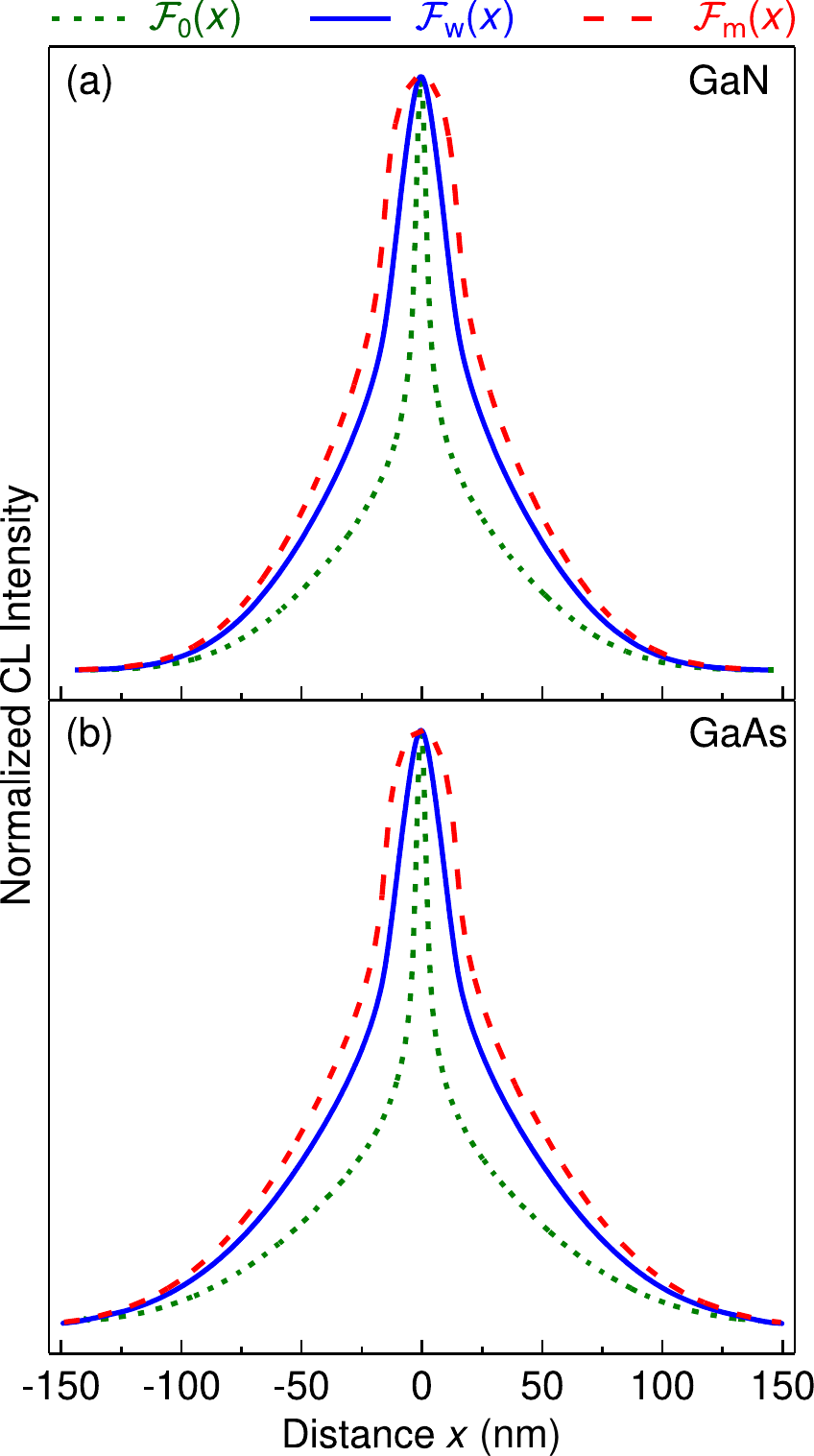}
    \caption{Simulated CL profiles obtained from single QWs embedded in (a) GaN  and (b) GaAs according to the experiment depicted in Fig.~\ref{fig1} and the three cases (i)--(iii) discussed in the text for $V=5$~kV. The center of the QW is situated at $x=0$. The dotted lines show the profiles in case (i), which are given by the energy loss distribution of the incident electron beam $\mathcal{F}_{0}(x)=\bar{Q}(x)$ calculated by \texttt{CASINO} \cite{drouin_2007} and integrated over the two coordinates $y$ and $z$ by Eq.~(\ref{eq:1}). The solid lines depict the CL profile in case (ii) taking into account the diffusion of carriers generated within the barriers (barrier width $b=15$~nm) towards the well and matrix material [$\mathcal{F}_{w}(x)$].  For the dashed lines representing the profiles in case (iii), we assume that all carriers generated within the barriers are captured by the well [$\mathcal{F}_{m}(x)$].} 
    \label{fig2} 
\end{figure}

We will first examine the influence of these different assumptions on the line profile for both GaN and GaAs. Let us assume that the initial distribution of carriers $Q(x,y,z)$ generated by the electron beam is given by the energy loss distribution of the incident electrons. This loss distribution is calculated using \texttt{CASINO} \cite{drouin_2007}, considering an In$_{0.16}$Ga$_{0.84}$N/Al$_{0.11}$Ga$_{0.89}$N and a GaAs/Al$_{0.4}$Ga$_{0.6}$As single QW with a width $2w=3$ and 7~nm, respectively, clad by barriers with a width $b=15$~nm. The acceleration voltage \textit{V} of the electron beam and the beam diameter are set to 5~kV and 5~nm, respectively \footnote{In the \texttt{CASINO} simulations, the density was set to 6.1~g/cm$^3$ for GaN and 5.3~g/cm$^3$ for GaAs. The default physical models were chosen, i.\,e., `Mott by interpolation' for the total and partial cross sections, `Casnati' for the effective ionisation potential, and `Joy and Luo' for the ionisation potential. The random number generator by `Press' and the direction cosines by `Drouin' were used.}. Figures~\ref{fig2}(a) and \ref{fig2}(b) show the simulated intensity profiles of the CL of the GaN- and GaAs-based QWs, respectively, considering the three different cases discussed above.  The profiles resulting from the assumption in case (i) are clearly narrower than those obtained by considering the contribution of the barriers in cases (ii) and (iii). Evidently, the impact of carriers excited in the barriers and transferred to the well is significant, and must not be ignored. However, the profiles taking into account the contributions from the barriers according to cases (ii) or (iii) differ only marginally. 

\section{Experiment}
\label{sec:experiment} 

For the experimental determination of the lateral extent of the CL generation volume, i.\,e., of $\mathcal{F}(x)$ in GaN and GaAs, we synthesized single QW structures for both materials systems by molecular beam epitaxy.  The former consists of a 3-nm-thick In$_{0.16}$Ga$_{0.84}$N well clad by 15-nm-thick Al$_{0.11}$Ga$_{0.89}$N barriers, embedded in a 1.3-\textmu m-thick GaN layer. This structure was synthesized on top of a GaN(0001) template prepared by metal-organic chemical vapor deposition on an Al$_2$O$_3$(0001) substrate. The sample has a threading dislocation density of $5\times10^8$~cm$^{-2}$ as measured in CD3 \cite{lahnemann_2020}. The latter structure contains a 7-nm-thick GaAs single QW clad by 15-nm-thick Al$_{0.4}$Ga$_{0.6}$As barriers within a 3-\textmu m-thick GaAs layer grown on a GaAs(001) substrate. In either case, we have chosen the structural and compositional parameters of the samples such as to inhibit carrier transport into and out of the QW by both thermionic emission and tunneling (see Appendix \ref{Appendix:Thermionic}), and to simultaneously ensure a sufficient spectral separation between the CL signals from the QW and the matrix. 

\begin{figure} 
    \includegraphics[width=0.80\columnwidth]{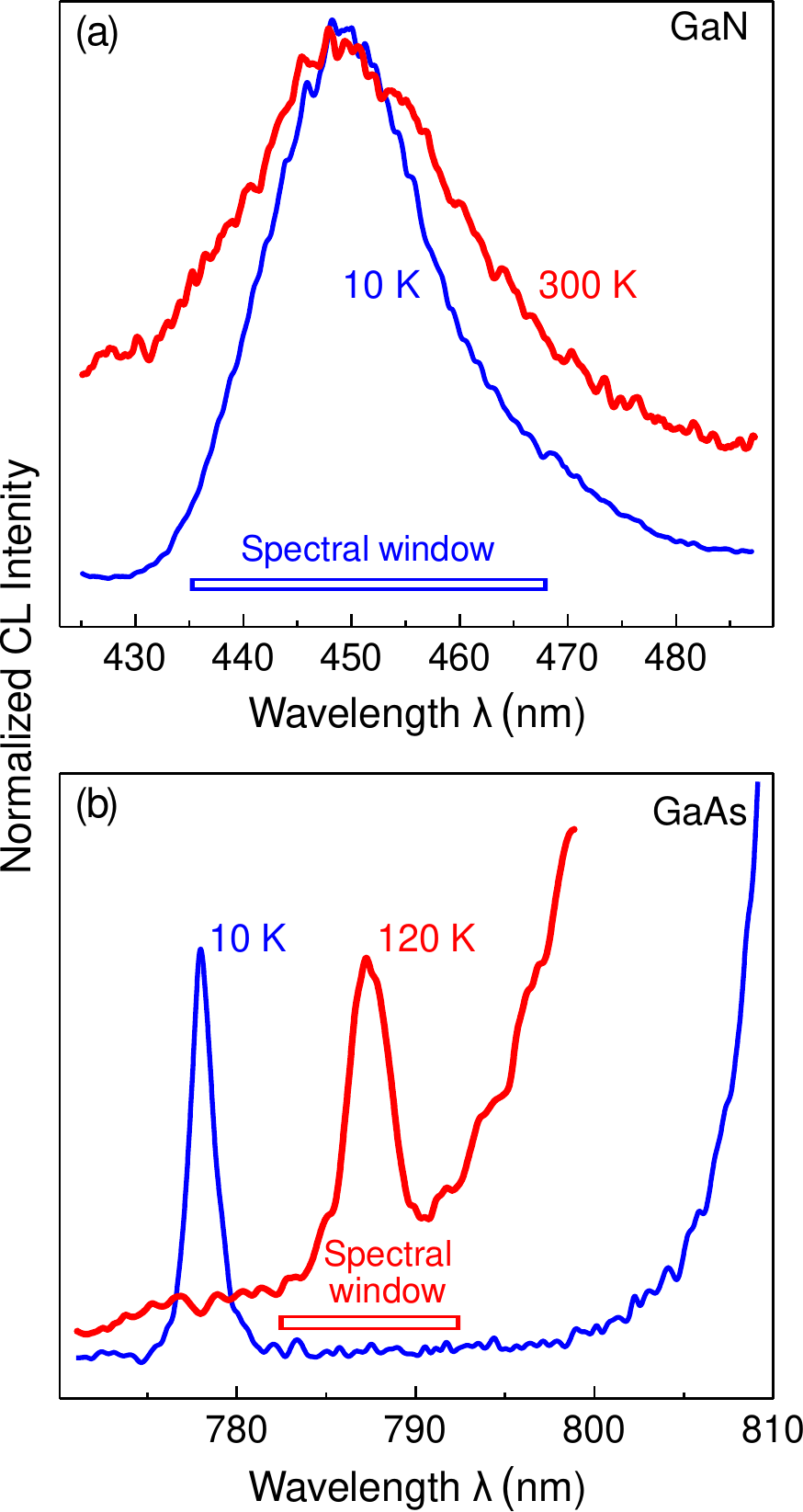}
    \caption{CL spectra acquired at the cross-sections of single QWs embedded in (a) GaN and (b) GaAs for $\textit{V}=5$~kV measured at low and elevated temperature.  The spectral windows used for recording monochromatic CL photon counting maps such as depicted in Fig.~\ref{fig4}(b) are indicated in the figure.} 
    \label{fig3}
\end{figure}

The CL investigations were performed using a Gatan mo\-noCL4 system and a He-cooling stage attached to a Zeiss Ultra55 SEM with a field emission gun. Cross-sectional specimens for performing CL linescans across the QW were obtained by cleaving the sample in air and introducing the freshly cleaved piece right after into the SEM vacuum chamber. CL intensity profiles across the QW were obtained from CL photon counting maps of the cleaved edge of the two samples under investigation. Figures~\ref{fig3}(a) and \ref{fig3}(b) show CL spectra of the GaN- and GaAs-related QWs, respectively, measured from a 1~\textmu m$^{2}$ region of the respective cross-section at low and elevated temperature.

\begin{figure} 
    \includegraphics[width=0.80\columnwidth]{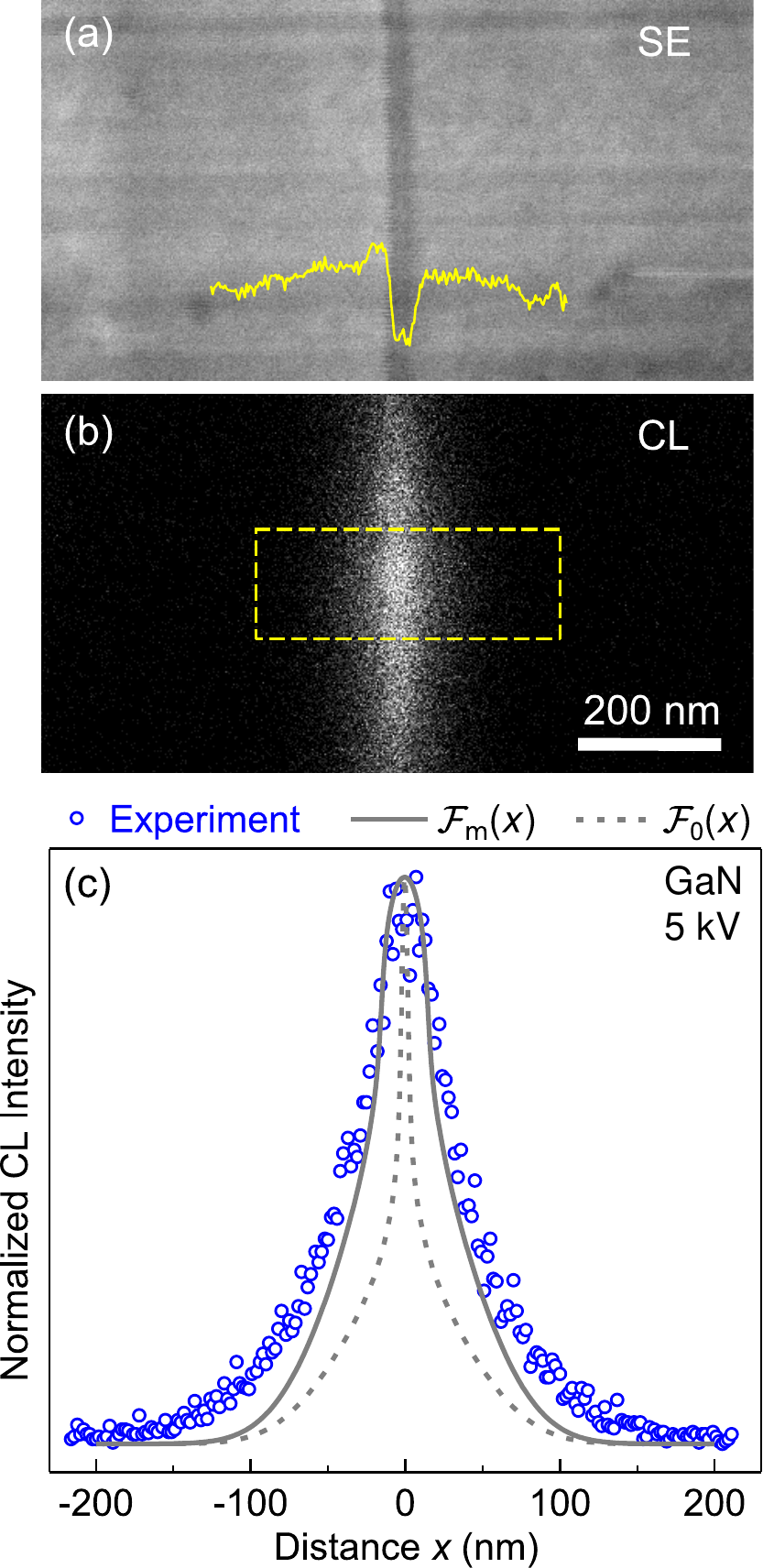}
    \caption{(a) SE micrograph and (b) monochromatic CL photon counting map at $\uplambda=450$~nm of the In$_{0.16}$Ga$_{0.84}$N/Al$_{0.11}$Ga$_{0.89}$N QW recorded at $T=120$~K and $V=5$~kV. The overlay on (a) shows a binned line profile across the center of the image to highlight the barrier and QW contrasts. In (b), the dashed rectangle indicates the window for the integration of the photon counts resulting in the experimental CL intensity profile. (c) Experimental CL intensity profile (symbols) extracted from the photon counting map in (b). The dashed and solid lines are the simulated profiles without [$\mathcal{F}_{0}(x)$] and with [$\mathcal{F}_{m}(x)$] taking into account carrier capture from the additional barriers, respectively.} 
    \label{fig4} 
\end{figure}

The spectral position of the (In,Ga)N/GaN QW CL band does not vary notably with varying temperature due to carrier localization effects.  In order to cover the whole spectrum while acquiring photon counting maps of the QW CL, the spectral window was set to 33~nm width as indicated in Fig.~\ref{fig3}(a). The CL spectrum of the GaAs/(Al,Ga)As QW depicted in Fig.~\ref{fig3}(b) shows a pronounced red-shift with increasing temperature and is superimposed by the emission spectrum of the GaAs matrix for $T > 140$~K. For the acquisition of monochromatic CL maps of the GaAs/(Al,Ga)As QW, we used a spectral window of about 10~nm width as indicated in Fig.~\ref{fig3}(b).  We recorded CL profiles of the GaN- and GaAs-related QW for electron beam energies between 3 and 15~keV, with the beam current varying between 0.3 and 0.7~nA. The experiments were performed at sample temperatures ranging from 10 to 300~K for the GaN- and 10 to 140~K for the GaAs-based QW.

\begin{figure*}
    \includegraphics[width=\textwidth]{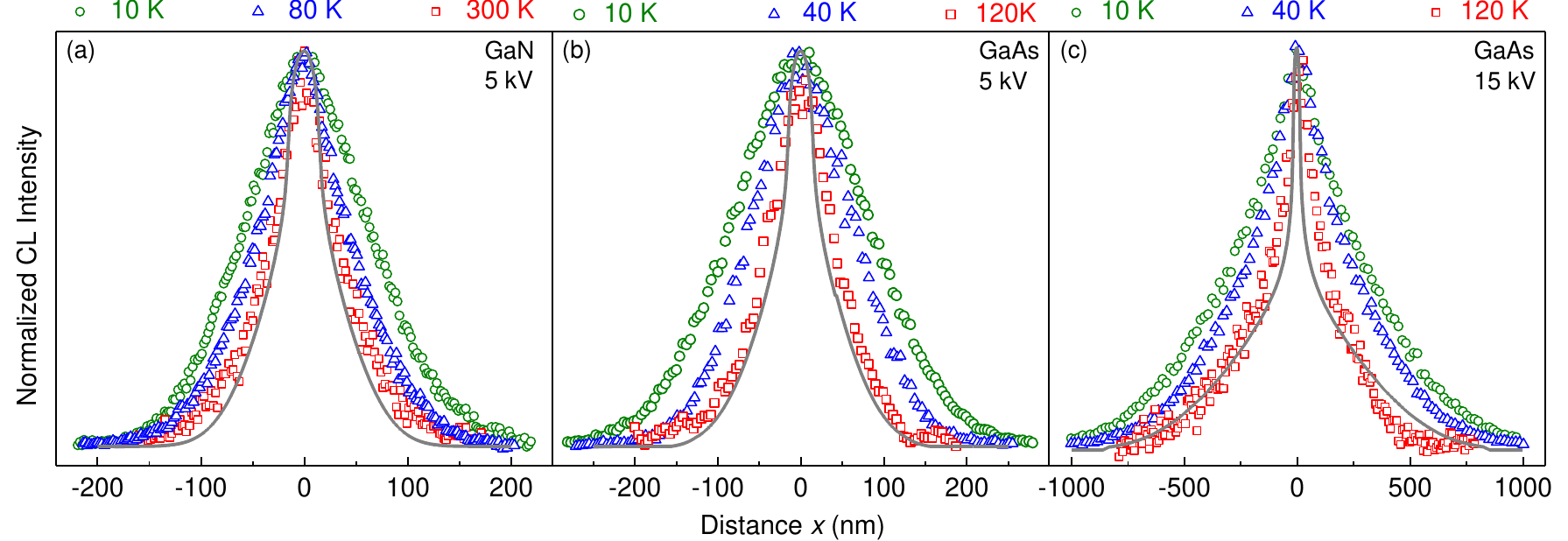}
    \caption{Experimental CL profiles across the (a) In$_{0.16}$Ga$_{0.84}$N/Al$_{0.11}$Ga$_{0.89}$N QW at $V=5$~kV and (b)/(c) GaAs/Al$_{0.4}$Ga$_{0.6}$As QW at $\textit{V}=5/15$~kV for different temperatures (symbols).  The lines represent the simulated profiles $\mathcal{F}_{m}(x)$ with the maximum width. Note the different $x$ scale used in (c).} 
    \label{fig5} 
\end{figure*}

\section{Results and discussion} 
\label{sec:results}

\subsection{Profiles of QW CL} \label{sec:CL profiles}

Figures~\ref{fig4}(a) and \ref{fig4}(b) display a secondary electron (SE) micrograph and a monochromatic CL photon counting map (central wavelength 450~nm), respectively, of the same cross-section region of the In$_{0.16}$Ga$_{0.84}$N/Al$_{0.11}$Ga$_{0.89}$N QW at 120~K and 5~kV. Due to the material contrast of the (Al,Ga)N barriers, the QW structure is clearly visible in the SE micrograph, where the high spatial resolution allows us to recognize even the 3-nm-thin QW in the center of the barrier-related dark stripe. The CL intensity map of Fig.~\ref{fig4}(b) reflects the scattering of the incident electrons into the QW structure and is consequently much broader than the SE micrograph of the QW. CL intensity profiles are obtained by integrating the counts within a 130-nm-wide stripe-like window intersecting the QW as indicated in Fig.~\ref{fig4}(b) by the dashed rectangle. Care was taken to choose areas with a homogeneous emission along the QW plane to avoid artifacts from dislocations or morphological defects induced during cleavage.

The symbols in Fig.~\ref{fig4}(c) represent the experimental CL profile thus obtained. The dashed and solid lines of Fig.~\ref{fig4}(c) are the simulated profiles based on \texttt{CASINO} without [$\mathcal{F}_{0}(x)$] and with [$\mathcal{F}_{m}(x)$] taking into account the profile broadening due to the presence of the barriers, respectively. This first example shows that even when accounting for the maximum broadening due to the barriers, the calculated profile is still slightly narrower than the experimental one.

Figures~\ref{fig5}(a) and \ref{fig5}(b) depict CL intensity profiles of the In$_{0.16}$Ga$_{0.84}$N/Al$_{0.11}$Ga$_{0.89}$N and GaAs/Al$_{0.4}$Ga$_{0.6}$As QWs (symbols), respectively, recorded for different temperatures at $\textit{V}=5$~kV.  The solid lines are the simulated profiles with the maximum width [cf.\ $\mathcal{F}_{m}(x)$ in Fig.~\ref{fig2}]. For the highest temperatures (300 and 120~K for GaN and GaAs, respectively), these simulated profiles are only marginally narrower then the experimental ones. With decreasing temperatures, however, the measured CL profiles become progressively broader for both GaN and GaAs. The same effect is observed also at significantly higher acceleration voltages, as exemplified for the GaAs QW at $\textit{V}=15$~kV in Fig.~\ref{fig5}(c). While the relative change of the profile width is comparable to that observed for $\textit{V}=5$~kV, the absolute one is larger.

The experimental data presented above demonstrate that the common perception of a temperature-independent generation volume in SEM, represented by the empirical energy loss distribution of the primary electrons as calculated, e.\,g., by \texttt{CASINO}, is incorrect as far as CL is concerned. Clearly, to account for our experimental findings, an additional temperature-dependent scattering mechanism has to be invoked that is not included in any of the presently available simulations of the generation volume in SEM. In the following, we will show that scattering of low-energy carriers by phonons is the most likely candidate for this mechanism.

\subsection{Electron-phonon scattering and CL profile width}
\label{sec:phonon}

In all applications of SEM, primary electrons are accelerated by voltages typically ranging from 1 to 30~kV. These primary electrons lose their energy once impinging onto the solid by a cascade of excitation processes, including high-energy events such as the excitation of core-level electrons, and low-energy events such as the excitation of plasmons that subsequently decay into hot electron-hole pairs, which in turn cool down by scattering with phonons.  In MC programs such as \texttt{CASINO}, the spatial distribution of these electron-hole pairs is represented by the total energy loss distribution of the impinging primary electrons calculated by empirical expressions \cite{joy_1989}, neglecting the energy relaxation and corresponding scattering events of the generated secondary carriers, and ignoring a potential dependence of these processes on temperature. This approximation is certainly justified for techniques for which low-energy electrons are irrelevant, such as for EDX. For CL, however, it has to be understood that efficient radiative recombination between electrons and holes only takes place when the respective carrier populations are close to thermalization with the lattice, i.\,e., when carriers have relaxed to their respective band edge in the vicinity of $\textbf{k}=0$. For energies below 20~eV, inelastic electron-electron scattering becomes rapidly inefficient with decreasing energy, and the energy loss rate of carriers is mainly determined by the Fröhlich interaction of electrons with longitudinal optical (LO) phonons \cite{dapor_2012}. For even lower energies, scattering with acoustic phonons contributes as well \cite{zhukov_2016}. In any case, the cooling of hot carriers is accompanied by an expansion of their spatial distribution, which can be understood as a (super)-diffusive process \cite{kozak_2015,najafi_2017}, but has to be distinguished from the normal diffusion of carriers in thermal equilibrium with the lattice. As a consequence of the inherent temperature dependence of electron-phonon scattering, we expect that the generation volume relevant for CL depends on temperature as well. 

\begin{figure}[b]
    \includegraphics[width=0.80\columnwidth]{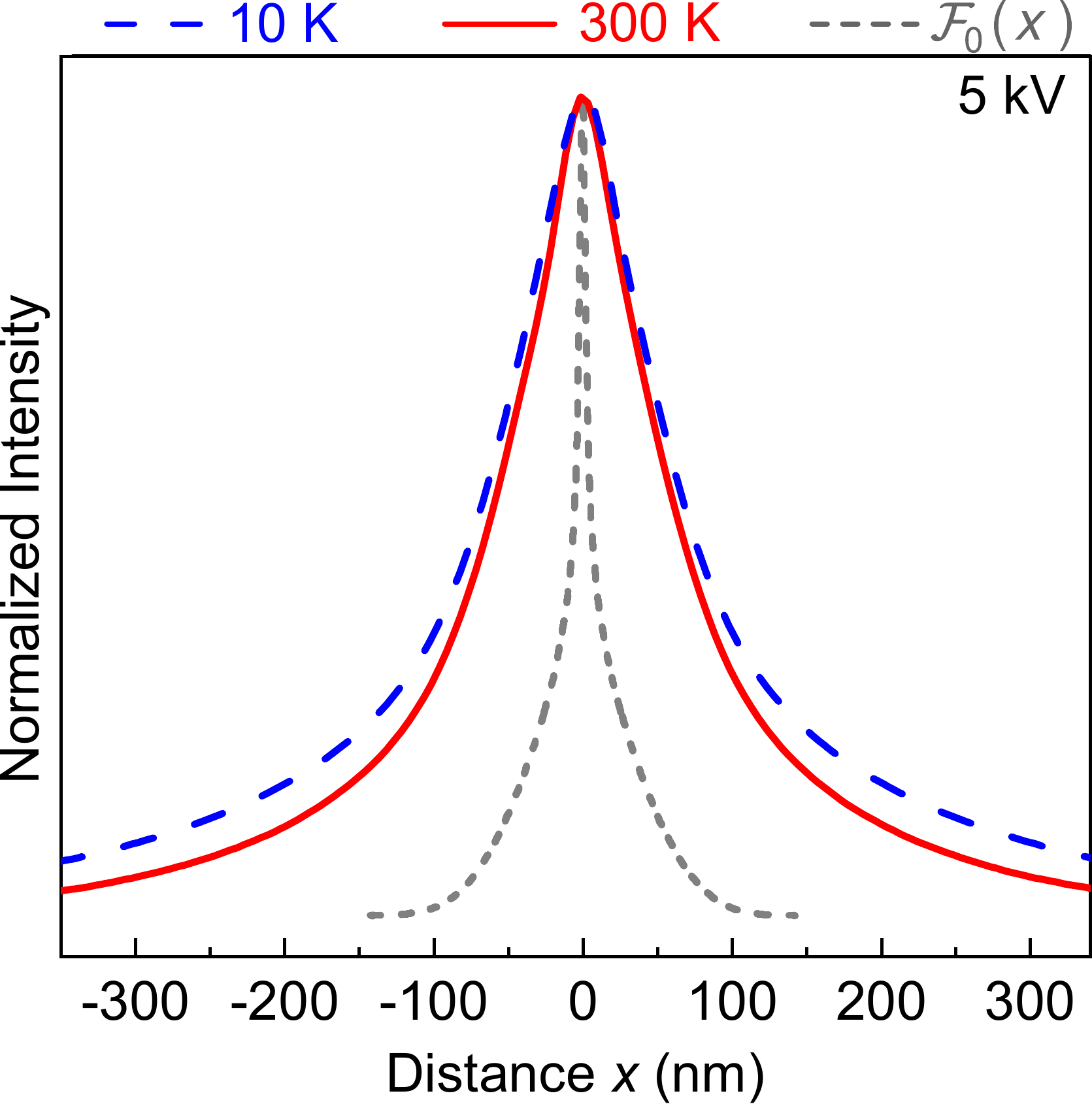} 
    \caption{Simulated lateral distributions of carriers after excitation of GaN at 5~kV and subsequent cooling at lattice temperatures of 10 (thick dashed line) and 300~K (thick solid line) explicitly taking into account the Fröhlich interaction as the sole inelastic scattering mechanism. The dotted curve depicts the initial lateral distribution of the energy loss $\mathcal{F}_{0}(x)$ calculated by \texttt{CASINO}.} 
    \label{fig6} 
\end{figure}

To see if this temperature dependence is qualitatively consistent with the effect observed experimentally, i.\,e., a widening of the profiles with decreasing temperature, we perform simple MC simulations as described in detail in Appendix \ref{Appendix:Electron-phonon}. These simulations are based on a toy model of carrier cooling that is reduced to the essence of energy relaxation via carrier-phonon scattering, but neglects details important for a quantitative description. Foremost, we assume that energy and momentum conservation are automatically fulfilled instead of considering the actual band structure and phonon dispersion of the semiconductor \cite{zhukov_2016}. Furthermore, we assume that both the effective mass and the effective phonon energy are constant, although both depend strongly on the electron excess energy \cite{Stanton_2001}. Finally, we neglect the Coulomb interaction between electrons and holes that becomes important at lower energies \cite{selbmann_1996}.

\begin{figure}
    \includegraphics[width=0.80\columnwidth]{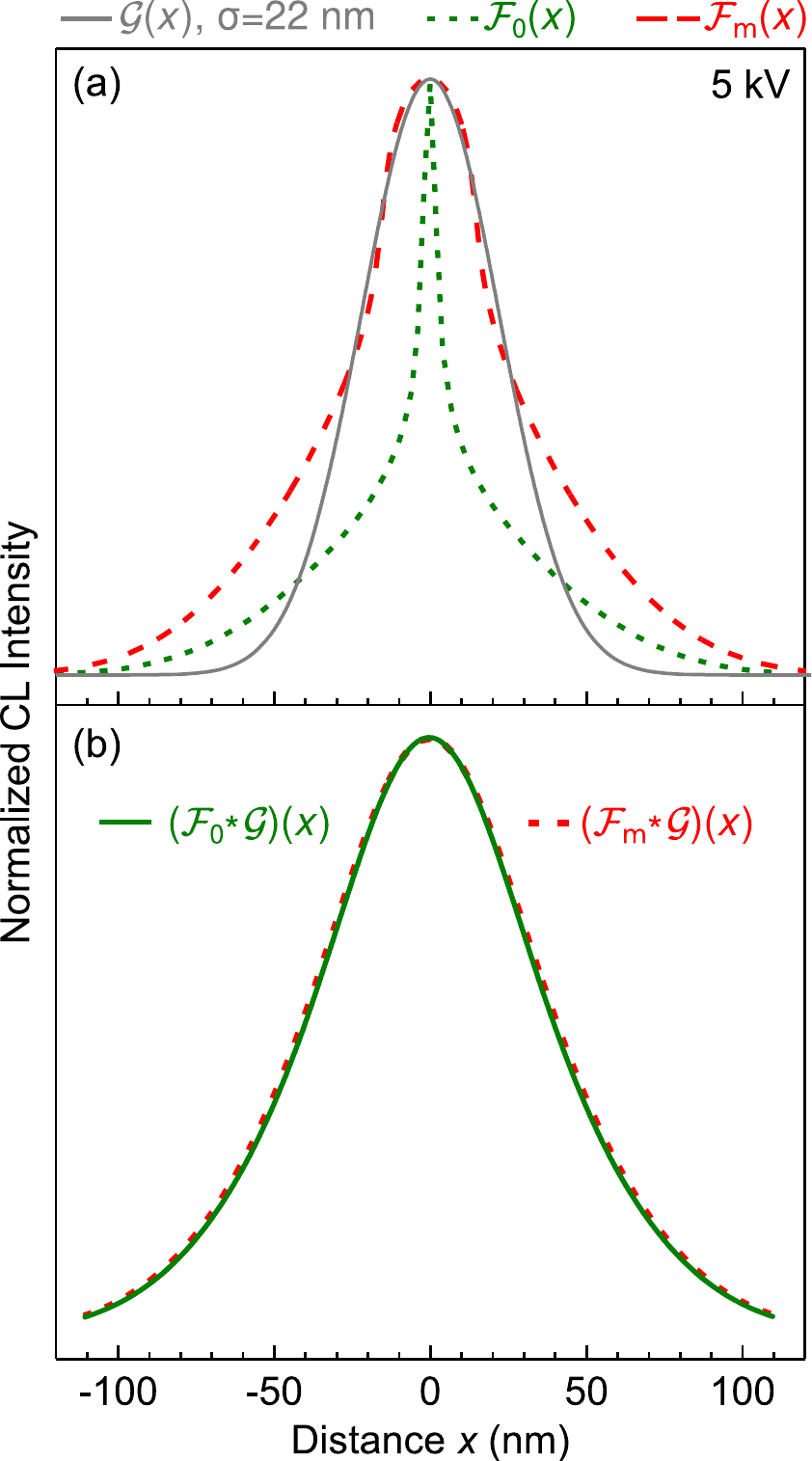} 
    \caption{(a) Intensity profiles given by the bare source $\mathcal{F}_{0}(x)$ (dotted line) and by $\mathcal{F}_{m}(x)$  taking into account the maximum broadening effect from the barriers (dashed line) for GaN and $V=5$~kV. The solid line depicts a Gaussian $\mathcal{G}(x)$ of width $\sigma = 22$~nm. (b) Convolution of $\mathcal{F}_{0}(x)$ and $\mathcal{F}_{m}(x)$ with $\mathcal{G}(x)$ resulting in the solid and dashed line, respectively.}
    \label{fig7} 
\end{figure}

We start with electrons with a three-dimensional spatial distribution as provided by \texttt{CASINO} for $V = 5$~kV, and a uniform energy distribution between an excess energy equal to the bandgap of GaN and the average phonon energy. We next allow these electrons to cool down via the Fröhlich interaction for lattice temperatures of 10 and 300~K and an effective phonon energy \cite{zhukov_2016} of 25~meV. The resulting three-dimensional distributions are reduced to a one-dimensional profile by integration. The profiles obtained for 10 and 300~K are displayed as dashed and solid lines in Fig.~\ref{fig6}, respectively. The dotted curve represents the corresponding lateral distribution of the energy loss of the primary electrons $\mathcal{F}_{0}(x)$ as calculated by \texttt{CASINO}. Evidently, our simple simulations qualitatively reproduce the main features observed experimentally: the cooling of hot carriers leads to a broadening of the energy loss profile, which is more pronounced at lower temperature. The simulations show that the reason underlying this effect is the increasing mean free length of carriers with decreasing temperature, which is determined almost exclusively by the lower probability of LO phonon emission.

\begin{figure*}
    \includegraphics[width=\textwidth]{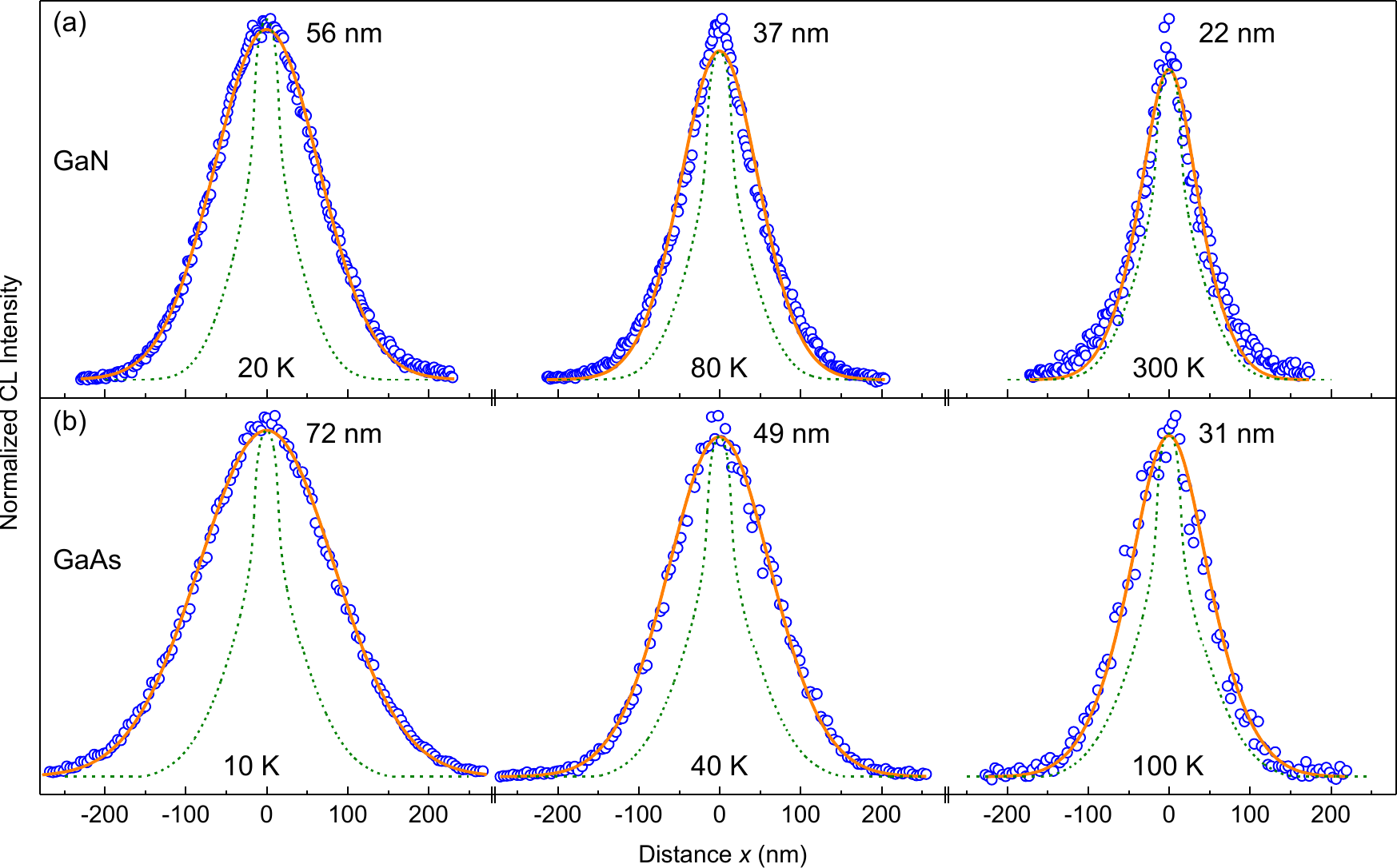}
    \caption{Experimental CL profiles (symbols) across the (a) In$_{0.16}$Ga$_{0.84}$N/Al$_{0.11}$Ga$_{0.89}$N QW and (b) GaAs/Al$_{0.4}$Ga$_{0.6}$As QW acquired at $V=5$~kV and different temperatures as indicated in the figure. The solid lines are fits to the data implemented by a convolution of the source $\mathcal{F}_{0}(x)$  with a Gaussian distribution $\mathcal{G}(x)$ of width $\sigma$ as indicated in the panels. The dotted curves represent $\mathcal{F}_{m}(x)$ with the maximum source broadening.} 
    \label{fig8} 
\end{figure*}

Instead of attempting to predict the temperature-dependent CL generation volume theoretically, which would be a challenging task in itself, we will simply extract its one-dimensional lateral component from our experimental data. In fact, the measured intensity profiles displayed in Fig.~\ref{fig5} contain the information about the temperature dependence of the CL generation volume. Understanding the broadening due to carrier thermalization as an independent and statistical process, we can phenomenologically describe the resulting profile as a convolution of $\mathcal{F}(x)$ and a Gaussian distribution $\mathcal{G}(x)$ with the standard deviation $\sigma$. Then, the flux of the carriers to the well can be written analogously to Eq.~(\ref{eq:7}), 
\begin{equation} 
\mathcal{\tilde{F}}(x)=\intop_{0}^{\infty}\bar{Q}(x-x')\tilde{F}(x')\,dx',\label{eq:2}
\end{equation} 
where the functions $\tilde{F}(x')$ are obtained by a convolution of any of the three functions $F(x')$ in Eq.~(\ref{eq:7}) with a Gaussian. The respective convolution integrals are shown to be given in terms of the error function in Appendix \ref{Appendix:Gaussian}.

To illustrate this approach, we compare a simulation based on the source $\mathcal{F}_{0}(x)$ with the corresponding one in which the maximum impact of barriers is taken into account via the distribution $\mathcal{F}_{m}(x)$.  Figure \ref{fig7}(a) shows these distributions together with a Gaussian distribution $\mathcal{G}(x)$ with $\sigma=22$~nm that we show below to be suitable for describing the room-temperature CL profile of the GaN-based QW [cf.\ Fig.~\ref{fig8}(a)]. Figure \ref{fig7}(b) displays the results of the convolution of these distributions with the Gaussian after normalization. Surprisingly, the shape of these convolutions is almost identical, and is unaffected by the presence of the barriers. Hence, we can extract the actual lateral carrier distribution by fitting experimental CL intensity profiles by a convolution of the energy loss distribution $\mathcal{F}_{0}(x)$ as obtained from \texttt{CASINO} and a Gaussian, with $\sigma$ being the sole fit parameter. In this way, it is straightforward to gather comprehensive information on the temperature- and voltage-dependent broadening of the source, which is essential for establishing an understanding of the mechanisms governing this broadening. 

\subsection{Temperature and voltage dependence of the CL profile width}
\label{sec:Lateral generation distribution}

Figures~\ref{fig8}(a) and \ref{fig8}(b) display experimental CL intensity profiles recorded at three different temperatures for both the GaN- and GaAs-based QWs, respectively. For both materials systems, the temperature-independent source $\mathcal{F}_{m}(x)$ severely underestimates the lateral extent of the CL generation volume even at the highest temperature and deviates increasingly from the measured extent of the CL generation volume with decreasing temperature. The convolution of $\mathcal{F}_{0}(x)$ with a Gaussian of width $\sigma(T)$ as proposed above as a phenomenological means to describe the profiles yields indeed satisfactory fits for all temperatures and for both materials systems.  

\begin{figure}[t] 
    \includegraphics[width=0.8\columnwidth]{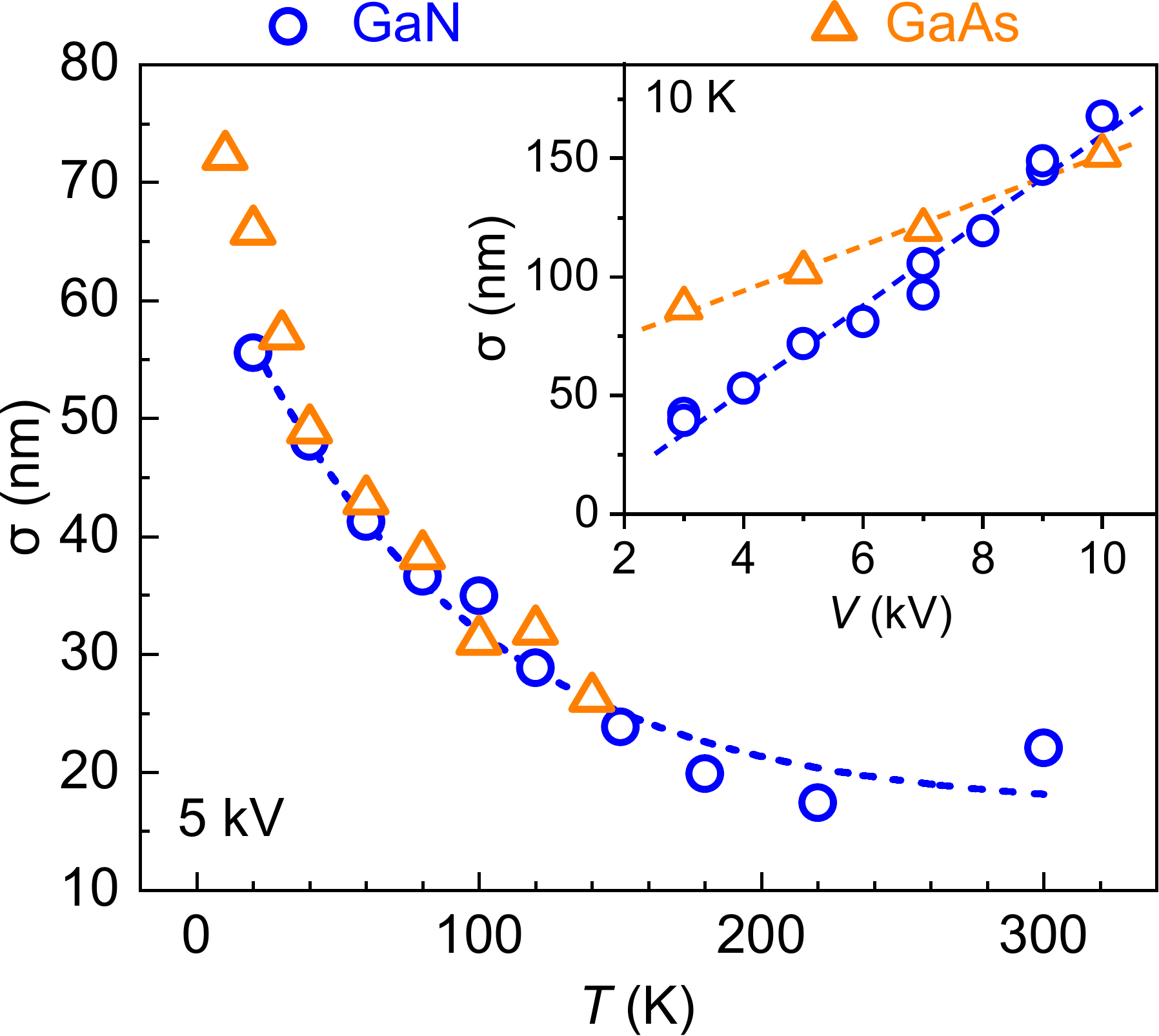}
    \caption{Standard deviation $\sigma$ of the Gaussian distribution $\mathcal{G}(x)$ obtained by fits of experimental CL profiles as shown exemplarily in Fig.~\ref{fig8}(a) and \ref{fig8}(b) as a function of \textit{T} for the single QWs embedded in GaN (circles) and GaAs (triangles). The dotted line is a guide to the eye. The inset shows the dependence of $\sigma$ on acceleration voltage for both the GaN- and the GaAs-based QW at 10~K.} 
    \label{fig9} 
\end{figure}

Figure~\ref{fig9} summarizes the results of the fits performed in this way  for the GaN- and GaAs-based QWs between 10 and 300~K and an acceleration voltage of 5~kV.  As discussed in detail above, the values of $\sigma$ represent the broadening of the source due to the cooling of hot carriers, and  the temperature dependence of this broadening is primarily related to the decreasing LO phonon emission rate with decreasing temperature. For the beam energy of 5~kV chosen for these experiments, the absolute values of $\sigma$ for the two samples are almost equal between 40 and 140~K, and become only slightly larger at lower temperatures for the QW embedded in GaAs. Overall, there seems to be no significant difference for the temperature-dependent broadening in the GaN- and GaAs-based QWs. However, this impression changes when examining profiles acquired at different acceleration voltages, as shown and discussed in detail in Appendix \ref{Appendix:Sigma}. In fact, below a characteristic temperature (which is different for GaN and GaAs), we observe a significant increase of $\sigma$ with  acceleration voltage. This effect is most pronounced at 10~K, at which $\sigma$ is found to increase linearly with \textit{V} for both the GaN- and the GaAs-based QW as shown in the inset in Fig.~\ref{fig9}. Interestingly, the slope of this increase is twice larger for the former as compared to the latter.

\begin{figure}
    \includegraphics[width=0.9\columnwidth]{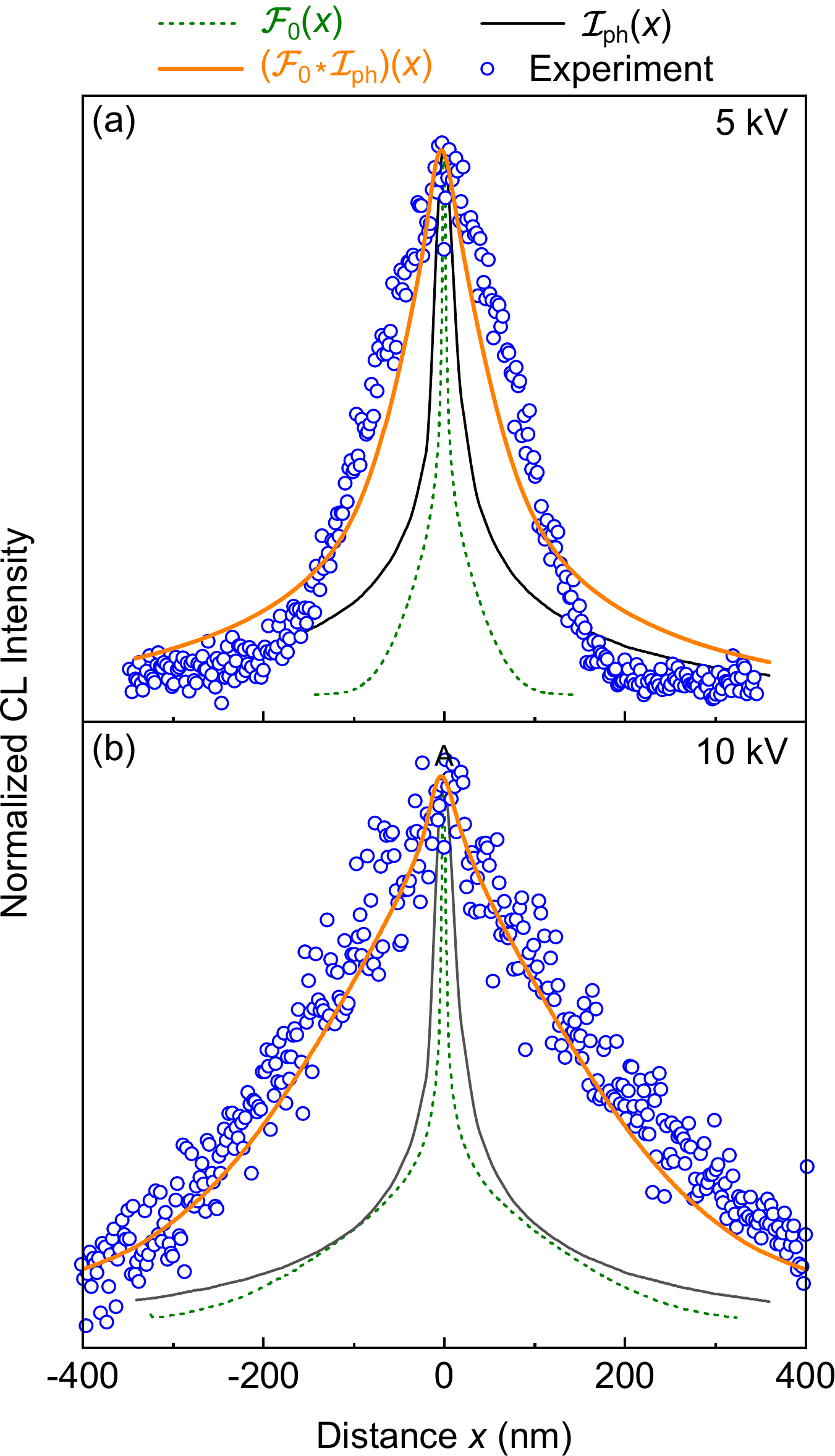}
    \caption{Comparison of the experimental CL profiles across the In$_{0.16}$Ga$_{0.84}$N/Al$_{0.11}$Ga$_{0.89}$N QW at (a) 5 and (b) 10~kV and 10~K with the simulated thermalized carrier distribution. Also shown are the source $\mathcal{F}_{0}(x)$ for the respective voltage and the distribution $\mathcal{I}_\textrm{ph}(x)$ resulting from electron-phonon scattering via the Fröhlich interaction. The thermalized carrier distribution $(\mathcal{F}_0*\mathcal{I}_\textrm{ph})(x)$ is obtained by a convolution of $\mathcal{F}_{0}(x)$ with $\mathcal{I}_\textrm{ph}(x)$.} 
    \label{fig10} 
\end{figure} 

At the first glance, the strong dependence of the temperature-dependent broadening of the profiles on the acceleration voltage or beam energy is puzzling, since the cooling of hot secondary electrons and holes by LO phonon emission is not expected to retain a memory of the energy of the primary electrons. The most obvious explanation seems to be a dependence of the carrier density on beam energy, since the cooling of hot carriers, or more precisely their energy loss rate, is known to depend on their density. In fact, experiments utilizing both continuous wave and pulsed laser excitation of semiconductors have shown that the energy loss rate of hot carriers decreases roughly linearly with increasing photogenerated carrier density \cite{leheny_1979,lugli_1987,lobentanzer_1987,leo_1988,marchetti_1989}. For GaAs, for example, the energy loss rate is reduced by one (two) orders of magnitude for a carrier density of $10^{17}$ ($10^{18}$)~cm$^{-3}$. This effect is caused by hot carriers cooling down via the emission of LO phonons, thus creating a nonequilibrium population of hot LO phonons \cite{shah_1970,shah_1986,lugli_1987} that in turn heats the carrier distribution by strongly increasing the probability of LO phonon absorption. As a consequence, carrier cooling slows down, causing a further quasi-diffusive broadening of the initial secondary carrier distribution.  

In the present experiments, four factors \footnote{Note that, in principle, surface recombination may affect the carrier density for low acceleration voltages, but can be safely ignored in the present case, since the \emph{M}-plane surface of GaN is known to exhibit a very low surface recombination velocity \cite{Corfdir_2014}.} determine the cathodogenerated carrier density: (i) the beam energy, with which the total number of carriers increases roughly linearly, (ii) the beam current, which usually increases sublinearly with the beam energy, (iii) the carrier lifetime, which depends strongly on temperature, but also on carrier density, and (iv) the generation volume, which depends strongly on beam energy, but also on temperature. For the case of GaN, for example, the first three of these factors result in an increase of the total number of carriers by a factor of about 30 from 3 to 10~kV, but this increase is overcompensated by the fourth factor, resulting in a carrier density that is actually even expected to \emph{decrease} with acceleration voltage, contrary to intuitive expectation.  In terms of absolute numbers, we estimate that the cathodogenerated carrier density for the GaN-based QW decreases from about $1 \times 10^{17}$ ~cm$^{-3}$ at 3~kV to $1 \times 10^{16}$ ~cm$^{-3}$ at 10~kV almost independent of temperature \footnote{The generation rate is determined according to \citet{Wu_1978} with the parameters given in Ref.~\onlinecite{Jahn_2003} and the generation volume approximated by a cylinder of diameter $\sigma$ and a height corresponding to 75\% of the \texttt{CASINO} energy loss distribution. The carrier density is then obtained with the carrier lifetimes measured by time-resolved photoluminescence spectroscopy [cf.\ Ref. \onlinecite{brandt_2020} (CD2) for the transients obtained for the barrierless (In,Ga)N/GaN QW].}. Obviously, this result rules out an increasing carrier density as an explanation for the voltage dependence of $\sigma$. 

The only other quantity that actually changes with acceleration voltage is the carrier \emph{distribution}. This fact is obvious when examining the lineshape of the energy loss profiles $\mathcal{F}_{0}(x)$ in GaN for acceleration voltages $V$ of 5 and 10~kV as displayed in Fig.~\ref{fig10}. While the central peak of $\mathcal{F}_{0}(x)$ hardly changes, the profile develops pronounced tails with increasing $V$. The profiles are thus not adequately described by their full-width at half maximum, but should rather be characterized by their integral breadth. The consequences of the heavy tails of $\mathcal{F}_{0}(x)$ can be elucidated by revisiting our MC simulations of carrier cooling (cf.\ Fig.~\ref{fig6}). These simulations were done by starting from the initial three-dimensional carrier distribution, and letting each electron cool down independent of each other from its initial energy by phonon scattering. The one-dimensional profiles shown in Fig.~\ref{fig6} are then obtained by integration along $y$ and $z$.

These profiles can also be derived in an alternative way that provides the key for understanding the voltage dependence of $\sigma$. The random walk of the carriers during their cooling is statistically the same for each point of the initial three-dimensional distribution of hot carriers. We can thus compute the random-walk distribution separately by performing simulations of carrier cooling for a point source. Both the initial three-dimensional distribution of hot carriers and the random-walk distribution can be reduced to one-dimensional profiles by integration over $y$ and $z$. The final one-dimensional carrier distribution after cooling is then simply given by a convolution of the one-dimensional distributions of the hot carriers $\mathcal{F}_{0}(x)$ and of their average random walk during the cooling process $\mathcal{I}_\textrm{ph}(x)$. The convolution $(\mathcal{F}_{0} * \mathcal{I}_\textrm{ph})(x)$ results in exactly the same profile as obtained in the three-dimensional simulation of the processes (cf.\ Fig.~\ref{fig6}).

Figures \ref{fig10}(a) and \ref{fig10}(b) compare the experimental CL profiles measured at 5 and 10~kV for the GaN-based QW with the convolutions of $\mathcal{F}_{0}(x)$ and $\mathcal{I}_\textrm{ph}(x)$. The profiles of the thermalized carrier distributions are obtained by convolutions of one and the same distribution $\mathcal{I}_\textrm{ph}(x)$, representing the thermalization of carriers, with the energy loss profile $\mathcal{F}_{0}(x)$ for the respective voltage. The resulting lineshapes in Fig.~\ref{fig10} do not exactly agree with the experimental ones, but the increase of the integral breadth of the profile is reproduced quantitatively. Since $\mathcal{I}_\textrm{ph}(x)$ does not depend on $V$, this increase can be fully attributed to the change of the lineshape of $\mathcal{F}_{0}(x)$ with acceleration voltage. We can understand this result by examining the peculiar lineshape of $\mathcal{F}_{0}(x)$ at 10~keV, which can be considered to be composed of a narrow central peak with heavy tails. These tails contain in fact the dominant fraction of the total number of carriers, so that the integral breadth of the profile is determined by the tails rather than by the narrow central peak. Correspondingly, the width of the convolution is governed by the tails, which explains the pronounced broadening of the CL profile with voltage. 

The increase of $\sigma$ as a function of voltage as shown in the inset of Fig.~\ref{fig9} can thus be understood easily. The distribution $\mathcal{I}_\textrm{ph}(x)$ representing the random walk of carriers thermalizing with the lattice is rather heavy-tailed (see Fig.~\ref{fig10}), very much in contrast to the Gaussian $\mathcal{G}(x)$ employed for our phenomenological lineshape fits (see Fig.~\ref{fig8}). The increase of $\sigma$ is a phenomenological way to reproduce the increase in width and integral breadth of the experimental CL profiles. The absolute values of $\sigma$ have no physical meaning, but the differences observed for $\sigma(T)$ in GaAs and GaN, as shown in Appendix \ref{Appendix:Sigma}, are meaningful and the result of the different material properties.

\section{Summary and Conclusions}
\label{sec:summary}

Our study has shown that the paradigm of a temperature-independent generation volume applicable to all SEM-based techniques needs to be revised. While MC simulations based on empirical energy loss expressions such as that implemented in \texttt{CASINO} describe the generation volume relevant for EDX very well, this generation volume is not the one relevant for CL. In fact, we have shown that the latter depends strongly on temperature, with a broadening that increases with decreasing temperature. This effect is understood when considering that radiative recombination of charge carriers in semiconductors takes place between thermalized electron-hole populations in the vicinity of $\mathbf{k} = 0$. The mean free path of carriers relaxing to their respective band edges is controlled by carrier-phonon scattering, and thus increases with decreasing temperature. In addition, we have found this phenomenon to be more pronounced for higher beam energies, which we have shown to be a consequence of the change of the shape of the initial carrier distribution.

In view of the considerable complexity of the energy relaxation of hot carriers in semiconductors, we have not attempted to develop a unified framework embracing the initial high-energy loss processes and the subsequent thermalization and cooling of hot carriers, but have opted for a phenomenological approach capable of approximating the temperature- and voltage-dependent lateral generation profiles reasonably well. We have shown that this goal can be achieved by convoluting the energy loss profile computed by \texttt{CASINO} with a Gaussian of variable standard deviation. We will use this methodology in the companion papers CD2 and CD3 for a reliable experimental determination of the temperature-dependent diffusivity in GaN.

\begin{acknowledgments} 
The authors are indebted to Achim Trampert for a critical reading of the manuscript and stimulating discussions. Special thanks are due to Holger Grahn, Lutz Geelhaar and Henning Riechert for their continuous encouragement and support. K.~K.~S.\ and A.~E.~K.\ acknowledge funding from the Russian Science Foundation under grant N 19-11-00019.
\end{acknowledgments}

\appendix %

\section{Thermionic emission and tunneling}
\label{Appendix:Thermionic}

Prior to any quantitative estimates, let us recall the specific conditions of the present experiments. Thermionic emission and tunneling are usually unipolar transport processes taking place in the presence of an electric field. In the present case, there is no net electric field across the barriers, and transport must be bipolar to affect the CL intensity. In other words, the probability for the combined charge transfer process depends on the product of the individual probabilities, i.\,e., essentially the sum of barrier heights in conduction and valence bands in the case of thermionic emission \cite{Botha_1994}, and the sum of the barrier widths for tunneling.    

Furthermore, let us consider the impact expected on the experimental CL profiles would either thermionic emission or tunneling compete with or even dominate over carrier recombination. The thermionic emission rate increases exponentially with temperature, and we would thus expect a broadening of the profiles with increasing temperature, contrary to what we observe experimentally. Tunneling is essentially temperature independent, and would thus give rise to profiles with constant width, which is again not in agreement with the experiment. In other words, our experimental results alone indicate that neither of these processes has an important impact for the CL profiles of the structures under investigation.     

The quantitative treatment of the transport of carriers across potential barriers in semiconductor heterostructures is a subject of considerable complexity, regardless of whether this transport is of classical or quantum-mechanical origin \cite{Schroeder_1994}. However, for an order-of-magnitude estimate, simplified treatments are feasible. We follow the consideration of thermionic emission and tunneling in QW structures as given by \citet{Schneider_1988} and \citet{Zou_1992}, respectively. In either of these approaches, the authors derive an escape time of electrons or holes from the QW surrounded by barriers with finite width and height.

For thermionic emission, the escape time is given by 
\begin{equation}
\frac{1}{\tau_{{\uparrow},i}} \approx \frac{\upsilon_{\text{th},i}}{4 w} \exp{(-\Delta E_i/k_B T)}
\label{eq:thermionic}
\end{equation}
with the thermal velocity $\upsilon_{\text{th},i} = (2 k_B T/\pi m^*_i)^{1/2}$, where $k_B$ is the Boltzmann constant, $T$ is the temperature, $m^*_i$ the effective mass of electrons or holes ($i= e, h$), and $\Delta E_i$ the potential barrier between the confined electron or hole state ($i= e, h$) in the QW and the band edge of the barrier.

For tunneling, the escape time reads
\begin{equation}
\frac{1}{\tau_{{\rightarrow},i}} \approx \frac{m^*_i \alpha^2 c^2}{\hbar}\frac{4 k_i K_i}{k_i^2+K_i^2} \exp{(-2 b_i^* K_i)}
\label{eq:tunneling}
\end{equation}
with $\alpha=e^2/(4 \pi \varepsilon_s \varepsilon_0 \hbar c)$, where $\varepsilon_s$ and $\varepsilon_0$ are the relative static and the vacuum permittivity, respectively, $\hbar$ is the reduced Planck constant, and $c$ the velocity of light. The wavevector $k_i = (2 m^*_i E_i/\hbar^2)^{1/2}$ applies to carriers residing in the QW and $K_i = [2 m^*_i (\Delta E_i - E_i)/\hbar^2]^{1/2}$ to those crossing the barrier. The effective barrier width $b_i^*$ can be different in the conduction and valence bands in the presence of internal electrostatic fields. Note that we have, for simplicity, assumed equal masses in well and barrier. 

The effective escape time containing contributions from both thermionic emission and tunneling can be written as
\begin{equation}
\frac{1}{\tau_{e,i}} = \frac{1}{\tau_{{\uparrow},i}} + \frac{1}{\tau_{{\rightarrow},i}}.
\end{equation}

To estimate the time $\tau_{c,i}$ for the opposite process, i.\,e., carriers crossing the barrier toward the QW, which could potentially broaden the CL profiles, we make use of the fact that the carrier fluxes into and out of the QW obey the principle of detailed balance in equilibrium, i.\,e., $n_{M}/\tau_{c,i} = n_{QW}/\tau_{e,i}$. Here, $n_{M}$ and $n_{QW}$ are the cathodogenerated sheet carrier densities in the matrix and in the QW, respectively, which are obtained by integrating the estimated volume density \cite{Note3} over $\pm \sigma/2 $ in the matrix and over $\pm w$ in the QW. The probability $p$ of carrier injection into the QW by either thermionic emission or tunneling is given by 
\begin{equation}
p_i = \frac{1/\tau_{c,i}}{1/\tau_{c,i} + 1/\tau} 
\end{equation}
with the carrier lifetime $\tau$, and $i=e,h$ again indicating whether electrons or holes are involved in the charge transfer. Finally, the probability $P$ of a bipolar transfer of an electron-hole pair into the QW is given by
\begin{equation}
P = \prod_{i=e,h} p_i .
\end{equation}

The most essential parameters in Eqs.~(\ref{eq:thermionic}) and (\ref{eq:tunneling}) are the barrier heights $\Delta E_i$ and the barrier widths $b_i^*$. Neglecting, for the moment, any internal electrostatic fields or band bending, the barrier heights are simply determined by the respective band gap differences and the band offset ratio. Assuming 60:40 for the latter, we obtain barrier heights of 300/200 and 600/400~meV for the conduction/valence bands in the GaAs- and GaN-based QWs, respectively. Likewise, the barrier widths are simply determined by the thickness of the Al-containing layer, which is 15~nm in both cases. With these numbers, the fastest process for both samples is actually the capture of holes by thermionic emission, for which we obtain a capture time on the order of 1~\textmu s at the highest temperatures for which CL profiles were recorded (140~K for the GaAs- and 300~K for the GaN-based QW). The capture of electrons is slower by about one to two orders of magnitude. Compared to the carrier lifetimes at these temperatures, which were measured to be 500 and 15~ps in the GaAs- and GaN-based sample, respectively, the carrier capture times are very long, resulting in values of $P$ for bipolar charge transfer as small as $3 \times 10^{-11}$ for the GaAs-based and $1 \times 10^{-14}$ for the GaN-based QW. Note that the latter number is actually even an upper bound for the GaN-based structure, for which both the barrier height and the barrier width are significantly enhanced by the internal electrostatic fields induced by the polarization discontinuities at the GaN/Al$_{0.11}$Ga$_{0.89}$N and Al$_{0.11}$Ga$_{0.89}$N/In$_{0.16}$Ga$_{0.84}$N interfaces. 

\section{Carrier cooling via LO phonon scattering}
\label{Appendix:Electron-phonon}

We assume the energy loss distribution as computed by \texttt{CASINO} to represent the initial spatial distribution of electrons (or holes). For constructing a situation in which all secondary excitations have energies below the impact ionization threshold, so that a generation of further electron-hole pairs cannot occur, we assume that they are uniformly distributed between the phonon energy and 3.5~eV, with the latter value representing the band gap and corresponding roughly to the impact ionization threshold of GaN \cite{zhukov_2016}. To participate in radiative recombination with likewise thermalized holes, these electrons have to relax to the bottom of the conduction band. For simplicity, we ignore that the thermalization takes place in a semiconductor with a complex band structure and phonon dispersion, and assume instead that momentum conservation is automatically fulfilled, and that thermalization occurs by the interaction of electrons having a constant mass $m^* = m_0$ equal to that of free electrons with phonons of an (effective) \cite{zhukov_2016} energy of 25~meV. To follow the evolution of the energy and the position of each electron during its energy relaxation process, we perform MC simulations in the spirit of those described in Refs.~\cite{llacer_1969,dapor_2012,dapor_2017}.

The mean free path of the electron between electron-phonon collisions is $\ell=\upsilon/(W^{+}+W^{-})$ with the electron velocity $\upsilon=\left(2 E/m_0\right)^{1/2}$, and the rates for phonon absorption and emission
\begin{equation} 
W^{-}=\frac{n_\textrm{ph}}{a_{B}\kappa}\frac{E_\textrm{ph}}{\sqrt{2 E m^{*}}}\ln\left(\frac{\left[1+E_\textrm{ph}/E\right]^{1/2}+1}{\left[1+E_\textrm{ph}/E\right]^{1/2}-1}\right),
\label{eq:2-1} 
\end{equation}
and 
\begin{equation} 
W^{+}=\frac{n_\textrm{ph}+1}{a_{B}\kappa}\frac{E_\textrm{ph}}{\sqrt{2 E m^{*}}}\ln\left(\frac{1+\left[1-E_\textrm{ph}/E\right]^{1/2}}{1-\left[1-E_\textrm{ph}/E\right]^{1/2}}\right),
\label{eq:3}
\end{equation} 
respectively. Here, $E_\textrm{ph}=0.025$~eV is the effective phonon energy, $a_B$ is the Bohr radius, $n_\textrm{ph}=\left(e^{E_\textrm{ph}/k_{B}T}-1\right)^{-1}$ is the equilibrium phonon occupation number, and $\kappa=\left(1/\varepsilon_{\infty}-1/\varepsilon_{s}\right)^{-1}=11.8$ with the high frequency and the static relative permittivities $\varepsilon_{\infty}$ and $\varepsilon_{s}$, respectively. 

Finally, the angular distribution of electrons scattered by phonons is given by the probability of scattering between $\theta$ and $\theta+d\theta$ proportional to 
\begin{equation} 
\frac{\sqrt{E'}\sin\theta\,d\theta}{E+E'-2\sqrt{EE'}\cos\theta},
\label{eq:4} 
\end{equation}
where $E$ and $E'$ are the electron energies before and after the scattering event.

\begin{figure}[b]
    \includegraphics[width=0.85\columnwidth]{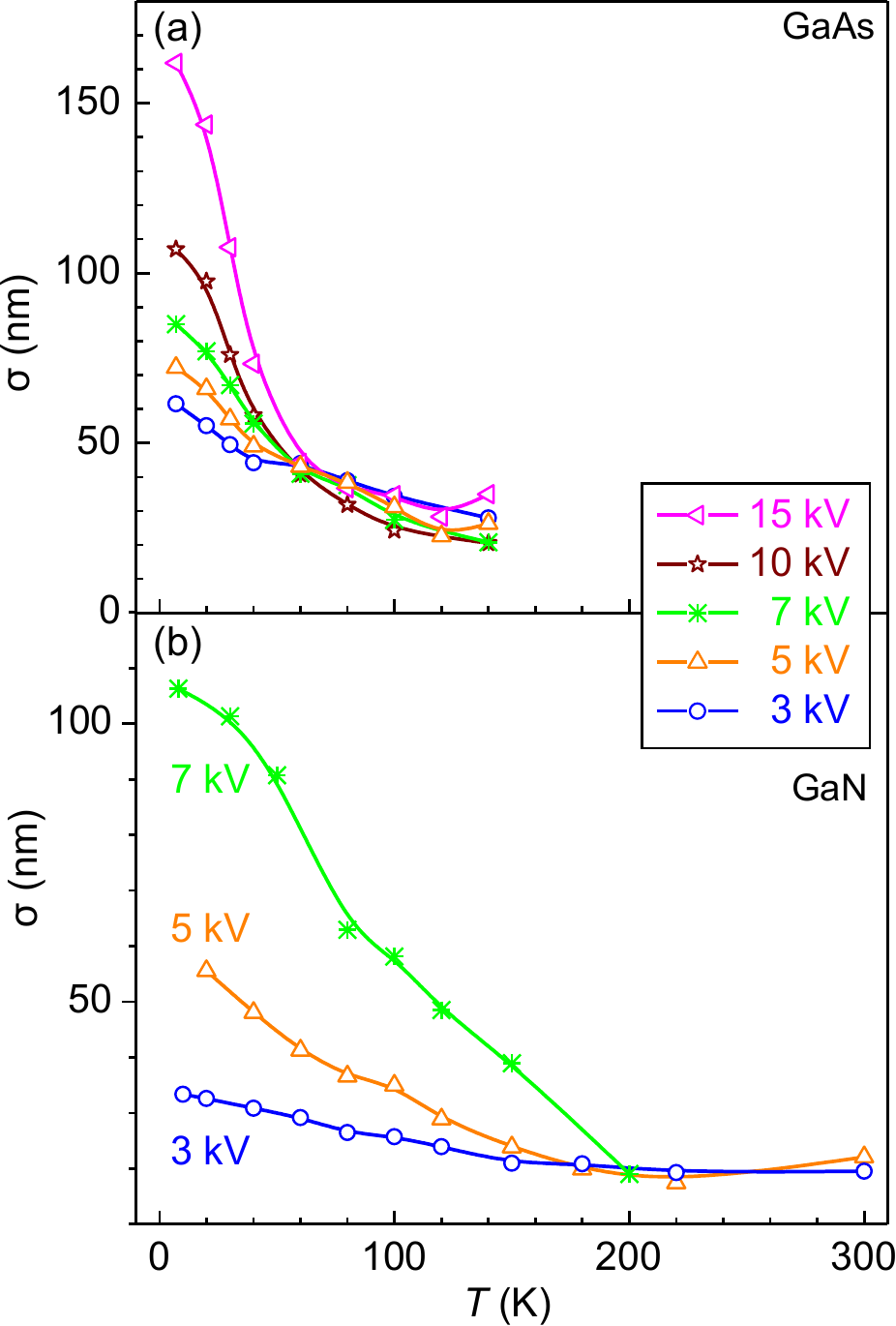}
    \caption{Standard deviation $\sigma$ of the Gaussian distribution $\mathcal{G}(x)$ obtained by fits of experimental CL profiles as a function of \textit{T} for various acceleration voltages of the primary electrons (a) for the GaAs- and (b) the GaN-based QW. The lines serve as guide to the eye.} 
    \label{fig11} 
\end{figure} 

\section{Model description of the lateral carrier distribution}
\label{Appendix:Gaussian}

Since a comprehensive description of the distribution of thermalized carriers produced by the electron beam is a highly complex problem, we model the source of carriers by assuming a Gaussian broadening of the energy loss distribution $Q(x,y,z)$ (that is calculated by CASINO), so that 
\begin{align} 
\tilde{Q}(x,y,z) & = \iintop_{-\infty}^{\infty}dx'\,dy'\,Q(x',y',z)\nonumber \\ &  \times\exp\left[-\frac{(x-x')^{2}+(y-y')^{2}}{2\sigma^{2}}\right].
\label{eq:s7} 
\end{align} 
In this model, we include convolutions over $x$ and $y$ but not over $z$, since the energy loss distribution $Q(x,y,z)$ is much broader in depth $z$ as compared with its width in the lateral directions $x,y$ and the actual standard deviation $\sigma$ required to match the experiment.

Our aim is to determine the total flux of the carriers $\mathcal{\tilde{F}}(x)$ to the quantum well as a function of the current position $x$ of the electron beam that scans along the surface. This flux is a result of the generation of carriers both directly in the well and in the barriers produced by a source $\tilde{Q}(x-x',y,z)$, where $x-x'$ are positions of the points in the source with respect to its center along the scan direction. As we have already discussed above, the $y$- and $z$-positions of the excitation points are not essential when calculating the total flux. We can consider a one-dimensional diffusion problem with the source 
\begin{equation} 
\tilde{\bar{Q}}(x-x')=\intop_{-\infty}^{\infty}dy\,\intop_{0}^{\infty}dz\,\tilde{Q}(x-x',y,z)
\label{eq:s1} 
\end{equation} 
and, collecting contributions from the carriers excited at all positions $x'$, represent the total flux as a convolution 
\begin{equation} 
\mathcal{\tilde{F}}(x)=\intop_{-\infty}^{\infty}dx'\,\tilde{\bar{Q}}(x-x')F(x').
\label{eq:s2} 
\end{equation} 
The function $F(x')$ is either equal to $F_w(x')$ [Eqs.~(\ref{eq:s4})] or $F_m(x')$ [Eqs.~(\ref{eq:s5})]. Using equations (\ref{eq:s7}), (\ref{eq:s1}), and (\ref{eq:s2}) and changing the sequence of integrations, we can write the total flux as a convolution 

\begin{equation} 
\mathcal{\tilde{F}}(x)=\intop_{-\infty}^{\infty}dx'\,\bar{Q}(x-x')\tilde{F}(x')
\label{eq:s8}
\end{equation} 
with 

\begin{equation} 
\tilde{F}(x)=\intop_{-\infty}^{\infty}dx'\,F(x')\exp\left[-\frac{(x-x')^{2}}{2\sigma^{2}}\right].
\label{eq:s9} 
\end{equation}

The integrals with the functions $F(x)$ given by Eqs.~(\ref{eq:s4}) and (\ref{eq:s5}) can be calculated analytically. For $F(x') =F_w(x')$, we get from Eq.~(\ref{eq:s4}) 

\begin{widetext}
\begin{align} 
\tilde{F}_{w}(x) & =  \frac{\sigma}{2b}\left\{ \sqrt{2\pi}\left[(b-x)\mathrm{erf}\left(\frac{b-x}{\sqrt{2}\sigma}\right)-2x\mathrm{erf}\left(\frac{x}{\sqrt{2}\sigma}\right)+(b+x)\mathrm{erf}\left(\frac{b+x}{\sqrt{2}\sigma}\right)\right]\right.\nonumber \\
& +\left.2\sigma\left[\exp\left(-\frac{(b-x)^{2}}{2\sigma^{2}}\right)-2\exp\left(-\frac{x^{2}}{2\sigma^{2}}\right)+\exp\left(-\frac{(b+x)^{2}}{2\sigma^{2}}\right)\right]\right\},
\label{eq:s11} 
\end{align} 
\end{widetext}
where $\mathrm{erf}(x)$ is the error function.

For $F(x') =F_m(x')$, we obtain from Eq.~(\ref{eq:s5}) 
\begin{equation} 
\tilde{F}_{m}(x)=\sigma\sqrt{\frac{\pi}{2}}\left[\mathrm{erf}\left(\frac{b-x}{\sqrt{2}\sigma}\right)+\mathrm{erf}\left(\frac{b+x}{\sqrt{2}\sigma}\right)\right].
\label{eq:s10}
\end{equation}

\section{Gaussian broadening for GaAs and GaN}
\label{Appendix:Sigma}

The following data are presented primarily for the benefit of those readers who would like to model the CL generation volume in GaN or GaAs for analyzing their own experiments. As described in detail in Sec.~\ref{sec:phonon}, this modeling consists of convoluting the one-dimensional energy loss profile $\mathcal{F}_0(x)$, which can be obtained directly from \texttt{CASINO}, with a Gaussian $\mathcal{G}(x)$ of width $\sigma$. 

Figures \ref{fig11}(a) and \ref{fig11}(b) show $\sigma(T)$ for the QWs embedded in GaAs and GaN, respectively, resulting from CL intensity profiles recorded with acceleration voltages between 3 and 15~kV. These plots demonstrate that the temperature dependencies of $\sigma$ for the GaAs- and GaN-based QWs are in fact quite different. In particular, three characteristic differences are noted: (i) At elevated temperature, $\sigma$ is essentially constant and amounts to about 30~nm for the GaAs-based and 20~nm for the GaN-based QW. (ii) At a characteristic temperature $T_C$, $\sigma$ starts to increase with decreasing temperature. $T_C$ corresponds to about 80~K for the GaAs-based and 150~K for the GaN-based QW independent of acceleration voltage. (iii) For $T < T_C$, $\sigma$ increases stronger with acceleration voltage for the GaN-based as compared to the GaAs-based QW.

We stress, once again, that the strong increase of $\sigma(T)$ with increasing $V$ actually reflects the change of the shape of the energy loss profile $\mathcal{F}_0(x)$ and has no physical meaning itself. However, this change of shape is not dramatically different for GaAs and GaN because of their similar densities (5.3 and 6.1~g/cm$^3$, respectively), and the characteristic differences in $\sigma$(\textit{T}) between GaAs and GaN are thus a result of their different material properties, particularly, the different LO phonon energy (36 and 92~meV, respectively) and effective strength of the electron-LO phonon coupling $\alpha$ (0.068 \cite{nash_1987} and 0.44 \cite{barker_1973}, respectively). The former is likely to be responsible for the difference in the onset of the broadening at lower temperatures, while the latter determines the dependence of the broadening on voltage. In fact, the reduction of the energy loss rate with temperature depends on the Fröhlich coupling constant: the stronger the coupling, the slower is the cooling. This relation has been demonstrated in a comparative study of the cooling of hot carriers in CdSe and GaAs, where the energy loss rate in CdSe was found to be reduced ten times more than in GaAs due to the stronger electron-LO phonon coupling in this more ionic II-VI semiconductor ($\alpha \approx 0.2$ for CdSe) \cite{prabhu_1995}.

\bibliography{references}

\begin{thebibliography}{50}%
\makeatletter
\providecommand \@ifxundefined [1]{%
 \@ifx{#1\undefined}
}%
\providecommand \@ifnum [1]{%
 \ifnum #1\expandafter \@firstoftwo
 \else \expandafter \@secondoftwo
 \fi
}%
\providecommand \@ifx [1]{%
 \ifx #1\expandafter \@firstoftwo
 \else \expandafter \@secondoftwo
 \fi
}%
\providecommand \natexlab [1]{#1}%
\providecommand \enquote  [1]{``#1''}%
\providecommand \bibnamefont  [1]{#1}%
\providecommand \bibfnamefont [1]{#1}%
\providecommand \citenamefont [1]{#1}%
\providecommand \href@noop [0]{\@secondoftwo}%
\providecommand \href [0]{\begingroup \@sanitize@url \@href}%
\providecommand \@href[1]{\@@startlink{#1}\@@href}%
\providecommand \@@href[1]{\endgroup#1\@@endlink}%
\providecommand \@sanitize@url [0]{\catcode `\\12\catcode `\$12\catcode
  `\&12\catcode `\#12\catcode `\^12\catcode `\_12\catcode `\%12\relax}%
\providecommand \@@startlink[1]{}%
\providecommand \@@endlink[0]{}%
\providecommand \url  [0]{\begingroup\@sanitize@url \@url }%
\providecommand \@url [1]{\endgroup\@href {#1}{\urlprefix }}%
\providecommand \urlprefix  [0]{URL }%
\providecommand \Eprint [0]{\href }%
\providecommand \doibase [0]{https://doi.org/}%
\providecommand \selectlanguage [0]{\@gobble}%
\providecommand \bibinfo  [0]{\@secondoftwo}%
\providecommand \bibfield  [0]{\@secondoftwo}%
\providecommand \translation [1]{[#1]}%
\providecommand \BibitemOpen [0]{}%
\providecommand \bibitemStop [0]{}%
\providecommand \bibitemNoStop [0]{.\EOS\space}%
\providecommand \EOS [0]{\spacefactor3000\relax}%
\providecommand \BibitemShut  [1]{\csname bibitem#1\endcsname}%
\let\auto@bib@innerbib\@empty
\bibitem [{\citenamefont {Reimer}(1998)}]{reimer_1998}%
  \BibitemOpen
  \bibfield  {author} {\bibinfo {author} {\bibfnamefont {L.}~\bibnamefont
  {Reimer}},\ }\href {https://doi.org/10.1007/978-3-540-38967-5} {\emph
  {\bibinfo {title} {Scanning Electron Microscopy}}},\ edited by\ \bibinfo
  {editor} {\bibfnamefont {P.~W.}\ \bibnamefont {Hawkes}}\ and\ \bibinfo
  {editor} {\bibfnamefont {H.~K.~V.}\ \bibnamefont {Lotsch}},\ \bibinfo
  {series} {Springer {{Series}} in {{Optical Sciences}}}, Vol.~\bibinfo
  {volume} {45}\ (\bibinfo  {publisher} {{Springer-Verlag}},\ \bibinfo
  {address} {{Berlin Heidelberg}},\ \bibinfo {year} {1998})\BibitemShut
  {NoStop}%
\bibitem [{\citenamefont {Goldstein}\ \emph {et~al.}(2003)\citenamefont
  {Goldstein}, \citenamefont {Newbury}, \citenamefont {Michael}, \citenamefont
  {Ritchie}, \citenamefont {Scott},\ and\ \citenamefont
  {Joy}}]{goldstein_2003}%
  \BibitemOpen
  \bibfield  {author} {\bibinfo {author} {\bibfnamefont {J.~I.}\ \bibnamefont
  {Goldstein}}, \bibinfo {author} {\bibfnamefont {D.~E.}\ \bibnamefont
  {Newbury}}, \bibinfo {author} {\bibfnamefont {J.~R.}\ \bibnamefont
  {Michael}}, \bibinfo {author} {\bibfnamefont {N.~W.}\ \bibnamefont
  {Ritchie}}, \bibinfo {author} {\bibfnamefont {J.~H.~J.}\ \bibnamefont
  {Scott}},\ and\ \bibinfo {author} {\bibfnamefont {D.~C.}\ \bibnamefont
  {Joy}},\ }\href {https://doi.org/10.1007/978-1-4615-0215-9} {\emph {\bibinfo
  {title} {Scanning Electron Microscopy and {{X}}-Ray Microanalysis}}}\
  (\bibinfo  {publisher} {{Springer Science + Business Media}},\ \bibinfo
  {address} {{New York}},\ \bibinfo {year} {2003})\BibitemShut {NoStop}%
\bibitem [{\citenamefont {Zhou}\ and\ \citenamefont {Wang}(2007)}]{zhou_2007}%
  \BibitemOpen
  \bibfield  {author} {\bibinfo {author} {\bibfnamefont {W.}~\bibnamefont
  {Zhou}}\ and\ \bibinfo {author} {\bibfnamefont {Z.~L.}\ \bibnamefont
  {Wang}},\ }\href {https://doi.org/10.1007/978-0-387-39620-0} {\emph {\bibinfo
  {title} {Scanning Microscopy for Nanotechnology: Techniques and
  Applications}}}\ (\bibinfo  {publisher} {{Springer}},\ \bibinfo {address}
  {{New York}},\ \bibinfo {year} {2007})\BibitemShut {NoStop}%
\bibitem [{\citenamefont {Coenen}\ and\ \citenamefont
  {Haegel}(2017)}]{coenen_2017}%
  \BibitemOpen
  \bibfield  {author} {\bibinfo {author} {\bibfnamefont {T.}~\bibnamefont
  {Coenen}}\ and\ \bibinfo {author} {\bibfnamefont {N.~M.}\ \bibnamefont
  {Haegel}},\ }\bibfield  {title} {\bibinfo {title} {Cathodoluminescence for
  the 21st century: Learning more from light},\ }\href
  {https://doi.org/10.1063/1.4985767} {\bibfield  {journal} {\bibinfo
  {journal} {Appl. Phys. Lett.}\ }\textbf {\bibinfo {volume} {4}},\ \bibinfo
  {pages} {031103} (\bibinfo {year} {2017})}\BibitemShut {NoStop}%
\bibitem [{\citenamefont {Lin}\ \emph {et~al.}(2017)\citenamefont {Lin},
  \citenamefont {Jahn}, \citenamefont {K{\"u}pers}, \citenamefont {Luna},
  \citenamefont {Lewis}, \citenamefont {Geelhaar},\ and\ \citenamefont
  {Brandt}}]{lin_2017}%
  \BibitemOpen
  \bibfield  {author} {\bibinfo {author} {\bibfnamefont {W.-H.}\ \bibnamefont
  {Lin}}, \bibinfo {author} {\bibfnamefont {U.}~\bibnamefont {Jahn}}, \bibinfo
  {author} {\bibfnamefont {H.}~\bibnamefont {K{\"u}pers}}, \bibinfo {author}
  {\bibfnamefont {E.}~\bibnamefont {Luna}}, \bibinfo {author} {\bibfnamefont
  {R.~B.}\ \bibnamefont {Lewis}}, \bibinfo {author} {\bibfnamefont
  {L.}~\bibnamefont {Geelhaar}},\ and\ \bibinfo {author} {\bibfnamefont
  {O.}~\bibnamefont {Brandt}},\ }\bibfield  {title} {\bibinfo {title}
  {Efficient methodology to correlate structural with optical properties of
  {{GaAs}} nanowires based on scanning electron microscopy},\ }\href
  {https://doi.org/10.1088/1361-6528/aa8394} {\bibfield  {journal} {\bibinfo
  {journal} {Nanotechnology}\ }\textbf {\bibinfo {volume} {28}},\ \bibinfo
  {pages} {415703} (\bibinfo {year} {2017})}\BibitemShut {NoStop}%
\bibitem [{\citenamefont {Edwards}\ and\ \citenamefont
  {Martin}(2011)}]{edwards_2011}%
  \BibitemOpen
  \bibfield  {author} {\bibinfo {author} {\bibfnamefont {P.~R.}\ \bibnamefont
  {Edwards}}\ and\ \bibinfo {author} {\bibfnamefont {R.~W.}\ \bibnamefont
  {Martin}},\ }\bibfield  {title} {\bibinfo {title} {Cathodoluminescence
  nano-characterization of semiconductors},\ }\href
  {https://doi.org/10.1088/0268-1242/26/6/064005} {\bibfield  {journal}
  {\bibinfo  {journal} {Semicond. Sci. Technol.}\ }\textbf {\bibinfo {volume}
  {26}},\ \bibinfo {pages} {064005} (\bibinfo {year} {2011})}\BibitemShut
  {NoStop}%
\bibitem [{\citenamefont {Everhart}\ and\ \citenamefont
  {Hoff}(1971)}]{everhart_1971}%
  \BibitemOpen
  \bibfield  {author} {\bibinfo {author} {\bibfnamefont {T.~E.}\ \bibnamefont
  {Everhart}}\ and\ \bibinfo {author} {\bibfnamefont {P.~H.}\ \bibnamefont
  {Hoff}},\ }\bibfield  {title} {\bibinfo {title} {Determination of kilovolt
  electron energy dissipation vs penetration distance in solid materials},\
  }\href {https://doi.org/10.1063/1.1660019} {\bibfield  {journal} {\bibinfo
  {journal} {J. Appl. Phys.}\ }\textbf {\bibinfo {volume} {42}},\ \bibinfo
  {pages} {5837} (\bibinfo {year} {1971})}\BibitemShut {NoStop}%
\bibitem [{\citenamefont {Fitting}\ \emph {et~al.}(1977)\citenamefont
  {Fitting}, \citenamefont {Glaefeke},\ and\ \citenamefont
  {Wild}}]{fitting_1977}%
  \BibitemOpen
  \bibfield  {author} {\bibinfo {author} {\bibfnamefont {H.-J.}\ \bibnamefont
  {Fitting}}, \bibinfo {author} {\bibfnamefont {H.}~\bibnamefont {Glaefeke}},\
  and\ \bibinfo {author} {\bibfnamefont {W.}~\bibnamefont {Wild}},\ }\bibfield
  {title} {\bibinfo {title} {Electron penetration and energy transfer in solid
  targets},\ }\href {https://doi.org/10.1002/pssa.2210430119} {\bibfield
  {journal} {\bibinfo  {journal} {Phys. Status Solidi B}\ }\textbf {\bibinfo
  {volume} {43}},\ \bibinfo {pages} {185} (\bibinfo {year} {1977})}\BibitemShut
  {NoStop}%
\bibitem [{\citenamefont {Donolato}(1981)}]{donolato_1981}%
  \BibitemOpen
  \bibfield  {author} {\bibinfo {author} {\bibfnamefont {C.}~\bibnamefont
  {Donolato}},\ }\bibfield  {title} {\bibinfo {title} {An analytical model of
  {{SEM}} and {{STEM}} charge collection images of dislocations in thin
  semiconductor layers: I. {{Minority}} carrier generation, diffusion, and
  collection},\ }\href {https://doi.org/10.1002/pssa.2210650231} {\bibfield
  {journal} {\bibinfo  {journal} {Phys. Status Solidi A}\ }\textbf {\bibinfo
  {volume} {65}},\ \bibinfo {pages} {649} (\bibinfo {year} {1981})}\BibitemShut
  {NoStop}%
\bibitem [{\citenamefont {Oelgart}\ and\ \citenamefont
  {Werner}(1984)}]{oelgart_1984}%
  \BibitemOpen
  \bibfield  {author} {\bibinfo {author} {\bibfnamefont {G.}~\bibnamefont
  {Oelgart}}\ and\ \bibinfo {author} {\bibfnamefont {U.}~\bibnamefont
  {Werner}},\ }\bibfield  {title} {\bibinfo {title} {Kilovolt electron energy
  loss distribution in {{GaAsP}}},\ }\href
  {https://doi.org/10.1002/pssa.2210850125} {\bibfield  {journal} {\bibinfo
  {journal} {Phys. Status Solidi A}\ }\textbf {\bibinfo {volume} {85}},\
  \bibinfo {pages} {205} (\bibinfo {year} {1984})}\BibitemShut {NoStop}%
\bibitem [{\citenamefont {Werner}\ \emph {et~al.}(1988)\citenamefont {Werner},
  \citenamefont {Koch},\ and\ \citenamefont {Oelgart}}]{werner_1988}%
  \BibitemOpen
  \bibfield  {author} {\bibinfo {author} {\bibfnamefont {U.}~\bibnamefont
  {Werner}}, \bibinfo {author} {\bibfnamefont {F.}~\bibnamefont {Koch}},\ and\
  \bibinfo {author} {\bibfnamefont {G.}~\bibnamefont {Oelgart}},\ }\bibfield
  {title} {\bibinfo {title} {Kilovolt electron energy loss distribution in
  {{Si}}},\ }\href {https://doi.org/10.1088/0022-3727/21/1/017} {\bibfield
  {journal} {\bibinfo  {journal} {J. Phys. D: Appl. Phys.}\ }\textbf {\bibinfo
  {volume} {21}},\ \bibinfo {pages} {116} (\bibinfo {year} {1988})}\BibitemShut
  {NoStop}%
\bibitem [{\citenamefont {Akamatsu}\ \emph {et~al.}(1989)\citenamefont
  {Akamatsu}, \citenamefont {Henoc},\ and\ \citenamefont
  {Martins}}]{akamatsu_1989}%
  \BibitemOpen
  \bibfield  {author} {\bibinfo {author} {\bibfnamefont {B.}~\bibnamefont
  {Akamatsu}}, \bibinfo {author} {\bibfnamefont {P.}~\bibnamefont {Henoc}},\
  and\ \bibinfo {author} {\bibfnamefont {R.~B.}\ \bibnamefont {Martins}},\
  }\bibfield  {title} {\bibinfo {title} {Caract\'erisation des dispositifs
  opto-\'electroniques par microscopie \`a balayage},\ }\href@noop {}
  {\bibfield  {journal} {\bibinfo  {journal} {J. Microsc. Spectrosc.
  Electron.}\ }\textbf {\bibinfo {volume} {14}},\ \bibinfo {pages} {12a}
  (\bibinfo {year} {1989})}\BibitemShut {NoStop}%
\bibitem [{\citenamefont {Holt}\ and\ \citenamefont
  {Napchan}(1994)}]{holt_1994}%
  \BibitemOpen
  \bibfield  {author} {\bibinfo {author} {\bibfnamefont {D.~B.}\ \bibnamefont
  {Holt}}\ and\ \bibinfo {author} {\bibfnamefont {E.}~\bibnamefont {Napchan}},\
  }\bibfield  {title} {\bibinfo {title} {Quantitation of {{SEM EBIC}} and
  {{CL}} signals using {{Monte Carlo}} electron-trajectory simulations},\
  }\href {https://doi.org/10.1002/sca.4950160203} {\bibfield  {journal}
  {\bibinfo  {journal} {Scanning}\ }\textbf {\bibinfo {volume} {16}},\ \bibinfo
  {pages} {78} (\bibinfo {year} {1994})}\BibitemShut {NoStop}%
\bibitem [{\citenamefont {Hovington}\ \emph {et~al.}(1997)\citenamefont
  {Hovington}, \citenamefont {Drouin},\ and\ \citenamefont
  {Gauvin}}]{hovington_1997}%
  \BibitemOpen
  \bibfield  {author} {\bibinfo {author} {\bibfnamefont {P.}~\bibnamefont
  {Hovington}}, \bibinfo {author} {\bibfnamefont {D.}~\bibnamefont {Drouin}},\
  and\ \bibinfo {author} {\bibfnamefont {R.}~\bibnamefont {Gauvin}},\
  }\bibfield  {title} {\bibinfo {title} {{{CASINO}}: A new {{Monte Carlo}} code
  in {{C}} language for electron beam interaction\textemdash{{Part I}}:
  Description of the program},\ }\href {https://doi.org/10.1002/sca.4950190101}
  {\bibfield  {journal} {\bibinfo  {journal} {Scanning}\ }\textbf {\bibinfo
  {volume} {19}},\ \bibinfo {pages} {1} (\bibinfo {year} {1997})}\BibitemShut
  {NoStop}%
\bibitem [{\citenamefont {Drouin}\ \emph {et~al.}(1997)\citenamefont {Drouin},
  \citenamefont {Hovington},\ and\ \citenamefont {Gauvin}}]{drouin_1997}%
  \BibitemOpen
  \bibfield  {author} {\bibinfo {author} {\bibfnamefont {D.}~\bibnamefont
  {Drouin}}, \bibinfo {author} {\bibfnamefont {P.}~\bibnamefont {Hovington}},\
  and\ \bibinfo {author} {\bibfnamefont {R.}~\bibnamefont {Gauvin}},\
  }\bibfield  {title} {\bibinfo {title} {{{CASINO}}: A new {{Monte Carlo}} code
  in {{C}} language for electron beam interactions\textemdash{{Part II}}:
  Tabulated values of the {{Mott}} cross section},\ }\href
  {https://doi.org/10.1002/sca.4950190103} {\bibfield  {journal} {\bibinfo
  {journal} {Scanning}\ }\textbf {\bibinfo {volume} {19}},\ \bibinfo {pages}
  {20} (\bibinfo {year} {1997})}\BibitemShut {NoStop}%
\bibitem [{\citenamefont {Drouin}\ \emph {et~al.}(2007)\citenamefont {Drouin},
  \citenamefont {Couture}, \citenamefont {Joly}, \citenamefont {Tastet},
  \citenamefont {Aimez},\ and\ \citenamefont {Gauvin}}]{drouin_2007}%
  \BibitemOpen
  \bibfield  {author} {\bibinfo {author} {\bibfnamefont {D.}~\bibnamefont
  {Drouin}}, \bibinfo {author} {\bibfnamefont {A.~R.}\ \bibnamefont {Couture}},
  \bibinfo {author} {\bibfnamefont {D.}~\bibnamefont {Joly}}, \bibinfo {author}
  {\bibfnamefont {X.}~\bibnamefont {Tastet}}, \bibinfo {author} {\bibfnamefont
  {V.}~\bibnamefont {Aimez}},\ and\ \bibinfo {author} {\bibfnamefont
  {R.}~\bibnamefont {Gauvin}},\ }\bibfield  {title} {\bibinfo {title} {{{CASINO
  V2}}.42\textemdash{{A}} fast and easy-to-use modeling tool for scanning
  electron microscopy and microanalysis users},\ }\href
  {https://doi.org/10.1002/sca.20000} {\bibfield  {journal} {\bibinfo
  {journal} {Scanning}\ }\textbf {\bibinfo {volume} {29}},\ \bibinfo {pages}
  {92} (\bibinfo {year} {2007})}\BibitemShut {NoStop}%
\bibitem [{\citenamefont {Demers}\ \emph {et~al.}(2011)\citenamefont {Demers},
  \citenamefont {{Poirier-Demers}}, \citenamefont {Couture}, \citenamefont
  {Joly}, \citenamefont {Guilmain}, \citenamefont {{de Jonge}},\ and\
  \citenamefont {Drouin}}]{demers_2011}%
  \BibitemOpen
  \bibfield  {author} {\bibinfo {author} {\bibfnamefont {H.}~\bibnamefont
  {Demers}}, \bibinfo {author} {\bibfnamefont {N.}~\bibnamefont
  {{Poirier-Demers}}}, \bibinfo {author} {\bibfnamefont {A.~R.}\ \bibnamefont
  {Couture}}, \bibinfo {author} {\bibfnamefont {D.}~\bibnamefont {Joly}},
  \bibinfo {author} {\bibfnamefont {M.}~\bibnamefont {Guilmain}}, \bibinfo
  {author} {\bibfnamefont {N.}~\bibnamefont {{de Jonge}}},\ and\ \bibinfo
  {author} {\bibfnamefont {D.}~\bibnamefont {Drouin}},\ }\bibfield  {title}
  {\bibinfo {title} {Three-dimensional electron microscopy simulation with the
  {{CASINO Monte Carlo}} software},\ }\href {https://doi.org/10.1002/sca.20262}
  {\bibfield  {journal} {\bibinfo  {journal} {Scanning}\ }\textbf {\bibinfo
  {volume} {33}},\ \bibinfo {pages} {135} (\bibinfo {year} {2011})}\BibitemShut
  {NoStop}%
\bibitem [{\citenamefont {Boyes}(2000)}]{boyes_2000}%
  \BibitemOpen
  \bibfield  {author} {\bibinfo {author} {\bibfnamefont {E.~D.}\ \bibnamefont
  {Boyes}},\ }\bibfield  {title} {\bibinfo {title} {On low voltage scanning
  electron microscopy and chemical microanalysis},\ }\href
  {https://doi.org/10.1017/s1431927602000545} {\bibfield  {journal} {\bibinfo
  {journal} {Microsc. Microanal.}\ }\textbf {\bibinfo {volume} {6}},\ \bibinfo
  {pages} {307} (\bibinfo {year} {2000})}\BibitemShut {NoStop}%
\bibitem [{\citenamefont {Llacer}\ and\ \citenamefont
  {Garwin}(1969)}]{llacer_1969}%
  \BibitemOpen
  \bibfield  {author} {\bibinfo {author} {\bibfnamefont {J.}~\bibnamefont
  {Llacer}}\ and\ \bibinfo {author} {\bibfnamefont {E.~L.}\ \bibnamefont
  {Garwin}},\ }\bibfield  {title} {\bibinfo {title} {Electron-{{Phonon}}
  interaction in alkali halides. {{I}}. {{The}} transport of secondary
  electrons with energies between 0.25 and 7.5 {{eV}}},\ }\href
  {https://doi.org/10.1063/1.1658075} {\bibfield  {journal} {\bibinfo
  {journal} {J. Appl. Phys.}\ }\textbf {\bibinfo {volume} {40}},\ \bibinfo
  {pages} {2766} (\bibinfo {year} {1969})}\BibitemShut {NoStop}%
\bibitem [{\citenamefont {Dapor}(2012)}]{dapor_2012}%
  \BibitemOpen
  \bibfield  {author} {\bibinfo {author} {\bibfnamefont {M.}~\bibnamefont
  {Dapor}},\ }\bibfield  {title} {\bibinfo {title} {Monte {{Carlo}} simulation
  of secondary electron emission from dielectric targets},\ }\href
  {https://doi.org/10.1088/1742-6596/402/1/012003} {\bibfield  {journal}
  {\bibinfo  {journal} {J. Phys. Conf. Ser.}\ }\textbf {\bibinfo {volume}
  {402}},\ \bibinfo {pages} {012003} (\bibinfo {year} {2012})}\BibitemShut
  {NoStop}%
\bibitem [{\citenamefont {Dapor}(2017)}]{dapor_2017}%
  \BibitemOpen
  \bibfield  {author} {\bibinfo {author} {\bibfnamefont {M.}~\bibnamefont
  {Dapor}},\ }\href {https://doi.org/10.1007/978-3-319-47492-2} {\emph
  {\bibinfo {title} {Transport of Energetic Electrons in Solids}}},\ \bibinfo
  {edition} {2nd}\ ed.,\ \bibinfo {series} {Springer Tracts in Modern Physics},
  Vol.\ \bibinfo {volume} {999}\ (\bibinfo  {publisher} {{Springer
  International Publishing}},\ \bibinfo {address} {{Cham}},\ \bibinfo {year}
  {2017})\BibitemShut {NoStop}%
\bibitem [{\citenamefont {Bonard}\ \emph {et~al.}(1996)\citenamefont {Bonard},
  \citenamefont {Gani{\`e}re}, \citenamefont {Akamatsu}, \citenamefont
  {Ara{\'u}jo},\ and\ \citenamefont {Reinhart}}]{bonard_1996}%
  \BibitemOpen
  \bibfield  {author} {\bibinfo {author} {\bibfnamefont {J.-M.}\ \bibnamefont
  {Bonard}}, \bibinfo {author} {\bibfnamefont {J.-D.}\ \bibnamefont
  {Gani{\`e}re}}, \bibinfo {author} {\bibfnamefont {B.}~\bibnamefont
  {Akamatsu}}, \bibinfo {author} {\bibfnamefont {D.}~\bibnamefont
  {Ara{\'u}jo}},\ and\ \bibinfo {author} {\bibfnamefont {F.-K.}\ \bibnamefont
  {Reinhart}},\ }\bibfield  {title} {\bibinfo {title} {Cathodoluminescence
  study of the spatial distribution of electron-hole pairs generated by an
  electron beam in {{Al{\textsubscript{0.4}}Ga{\textsubscript{0.6}}As}}},\
  }\href {https://doi.org/10.1063/1.362560} {\bibfield  {journal} {\bibinfo
  {journal} {J. Appl. Phys.}\ }\textbf {\bibinfo {volume} {79}},\ \bibinfo
  {pages} {8693} (\bibinfo {year} {1996})}\BibitemShut {NoStop}%
\bibitem [{\citenamefont {Brandt}\ \emph {et~al.}(2020)\citenamefont {Brandt},
  \citenamefont {Kaganer}, \citenamefont {L{\"a}hnemann}, \citenamefont
  {Flissikowski}, \citenamefont {Pf{\"u}ller}, \citenamefont {Sabelfeld},
  \citenamefont {Kireeva}, \citenamefont {Ch{\`e}ze}, \citenamefont {Calarco},
  \citenamefont {Grahn},\ and\ \citenamefont {Jahn}}]{brandt_2020}%
  \BibitemOpen
  \bibfield  {author} {\bibinfo {author} {\bibfnamefont {O.}~\bibnamefont
  {Brandt}}, \bibinfo {author} {\bibfnamefont {V.~M.}\ \bibnamefont {Kaganer}},
  \bibinfo {author} {\bibfnamefont {J.}~\bibnamefont {L{\"a}hnemann}}, \bibinfo
  {author} {\bibfnamefont {T.}~\bibnamefont {Flissikowski}}, \bibinfo {author}
  {\bibfnamefont {C.}~\bibnamefont {Pf{\"u}ller}}, \bibinfo {author}
  {\bibfnamefont {K.~K.}\ \bibnamefont {Sabelfeld}}, \bibinfo {author}
  {\bibfnamefont {A.~E.}\ \bibnamefont {Kireeva}}, \bibinfo {author}
  {\bibfnamefont {C.}~\bibnamefont {Ch{\`e}ze}}, \bibinfo {author}
  {\bibfnamefont {R.}~\bibnamefont {Calarco}}, \bibinfo {author} {\bibfnamefont
  {H.~T.}\ \bibnamefont {Grahn}},\ and\ \bibinfo {author} {\bibfnamefont
  {U.}~\bibnamefont {Jahn}},\ }\bibfield  {title} {\bibinfo {title} {Carrier
  diffusion in {{GaN}} -- a cathodoluminescence study. {{II}}: Ambipolar vs.
  exciton diffusion},\ }\href {http://arxiv.org/abs/2009.13983} {\bibfield
  {journal} {\bibinfo  {journal} {arXiv:2009.13983 [cond-mat,
  physics:physics]}\ } (\bibinfo {year} {2020})}\BibitemShut {NoStop}%
\bibitem [{\citenamefont {L{\"a}hnemann}\ \emph {et~al.}(2020)\citenamefont
  {L{\"a}hnemann}, \citenamefont {Kaganer}, \citenamefont {Sabelfeld},
  \citenamefont {Kireeva}, \citenamefont {Jahn}, \citenamefont {Ch{\`e}ze},
  \citenamefont {Calarco},\ and\ \citenamefont {Brandt}}]{lahnemann_2020}%
  \BibitemOpen
  \bibfield  {author} {\bibinfo {author} {\bibfnamefont {J.}~\bibnamefont
  {L{\"a}hnemann}}, \bibinfo {author} {\bibfnamefont {V.~M.}\ \bibnamefont
  {Kaganer}}, \bibinfo {author} {\bibfnamefont {K.~K.}\ \bibnamefont
  {Sabelfeld}}, \bibinfo {author} {\bibfnamefont {A.~E.}\ \bibnamefont
  {Kireeva}}, \bibinfo {author} {\bibfnamefont {U.}~\bibnamefont {Jahn}},
  \bibinfo {author} {\bibfnamefont {C.}~\bibnamefont {Ch{\`e}ze}}, \bibinfo
  {author} {\bibfnamefont {R.}~\bibnamefont {Calarco}},\ and\ \bibinfo {author}
  {\bibfnamefont {O.}~\bibnamefont {Brandt}},\ }\bibfield  {title} {\bibinfo
  {title} {Carrier diffusion in {{GaN}} -- a cathodoluminescence study.
  {{III}}: Nature of nonradiative recombination at threading dislocations},\
  }\href {http://arxiv.org/abs/2009.14634} {\bibfield  {journal} {\bibinfo
  {journal} {arXiv:2009.14634 [cond-mat, physics:physics]}\ } (\bibinfo {year}
  {2020})}\BibitemShut {NoStop}%
\bibitem [{Note1()}]{Note1}%
  \BibitemOpen
  \bibinfo {note} {In the \protect \texttt {CASINO} simulations, the density
  was set to 6.1~g/cm$^3$ for GaN and 5.3~g/cm$^3$ for GaAs. The default
  physical models were chosen, i.\protect \,e., `Mott by interpolation' for the
  total and partial cross sections, `Casnati' for the effective ionisation
  potential, and `Joy and Luo' for the ionisation potential. The random number
  generator by `Press' and the direction cosines by `Drouin' were
  used.}\BibitemShut {Stop}%
\bibitem [{\citenamefont {Joy}\ and\ \citenamefont {Luo}(1989)}]{joy_1989}%
  \BibitemOpen
  \bibfield  {author} {\bibinfo {author} {\bibfnamefont {D.~C.}\ \bibnamefont
  {Joy}}\ and\ \bibinfo {author} {\bibfnamefont {S.}~\bibnamefont {Luo}},\
  }\bibfield  {title} {\bibinfo {title} {An empirical stopping power
  relationship for low-energy electrons},\ }\href
  {https://doi.org/10.1002/sca.4950110404} {\bibfield  {journal} {\bibinfo
  {journal} {Scanning}\ }\textbf {\bibinfo {volume} {11}},\ \bibinfo {pages}
  {176} (\bibinfo {year} {1989})}\BibitemShut {NoStop}%
\bibitem [{\citenamefont {Zhukov}\ \emph {et~al.}(2016)\citenamefont {Zhukov},
  \citenamefont {Tyuterev}, \citenamefont {Chulkov},\ and\ \citenamefont
  {Echenique}}]{zhukov_2016}%
  \BibitemOpen
  \bibfield  {author} {\bibinfo {author} {\bibfnamefont {V.~P.}\ \bibnamefont
  {Zhukov}}, \bibinfo {author} {\bibfnamefont {V.~G.}\ \bibnamefont
  {Tyuterev}}, \bibinfo {author} {\bibfnamefont {E.~V.}\ \bibnamefont
  {Chulkov}},\ and\ \bibinfo {author} {\bibfnamefont {P.~M.}\ \bibnamefont
  {Echenique}},\ }\bibfield  {title} {\bibinfo {title} {Electron-phonon
  relaxation and excited electron distribution in gallium nitride},\ }\href
  {https://doi.org/10.1063/1.4961874} {\bibfield  {journal} {\bibinfo
  {journal} {J. Appl. Phys.}\ }\textbf {\bibinfo {volume} {120}},\ \bibinfo
  {pages} {085708} (\bibinfo {year} {2016})}\BibitemShut {NoStop}%
\bibitem [{\citenamefont {Koz{\'a}k}\ \emph {et~al.}(2015)\citenamefont
  {Koz{\'a}k}, \citenamefont {Troj{\'a}nek},\ and\ \citenamefont
  {Mal{\'y}}}]{kozak_2015}%
  \BibitemOpen
  \bibfield  {author} {\bibinfo {author} {\bibfnamefont {M.}~\bibnamefont
  {Koz{\'a}k}}, \bibinfo {author} {\bibfnamefont {F.}~\bibnamefont
  {Troj{\'a}nek}},\ and\ \bibinfo {author} {\bibfnamefont {P.}~\bibnamefont
  {Mal{\'y}}},\ }\bibfield  {title} {\bibinfo {title} {Hot-carrier transport in
  diamond controlled by femtosecond laser pulses},\ }\href
  {https://doi.org/10.1088/1367-2630/17/5/053027} {\bibfield  {journal}
  {\bibinfo  {journal} {New J. Phys.}\ }\textbf {\bibinfo {volume} {17}},\
  \bibinfo {pages} {053027} (\bibinfo {year} {2015})}\BibitemShut {NoStop}%
\bibitem [{\citenamefont {Najafi}\ \emph {et~al.}(2017)\citenamefont {Najafi},
  \citenamefont {Ivanov}, \citenamefont {Zewail},\ and\ \citenamefont
  {Bernardi}}]{najafi_2017}%
  \BibitemOpen
  \bibfield  {author} {\bibinfo {author} {\bibfnamefont {E.}~\bibnamefont
  {Najafi}}, \bibinfo {author} {\bibfnamefont {V.}~\bibnamefont {Ivanov}},
  \bibinfo {author} {\bibfnamefont {A.}~\bibnamefont {Zewail}},\ and\ \bibinfo
  {author} {\bibfnamefont {M.}~\bibnamefont {Bernardi}},\ }\bibfield  {title}
  {\bibinfo {title} {Super-diffusion of excited carriers in semiconductors},\
  }\href {https://doi.org/10.1038/ncomms15177} {\bibfield  {journal} {\bibinfo
  {journal} {Nat. Commun.}\ }\textbf {\bibinfo {volume} {8}},\ \bibinfo {pages}
  {1} (\bibinfo {year} {2017})}\BibitemShut {NoStop}%
\bibitem [{\citenamefont {Stanton}\ \emph {et~al.}(2001)\citenamefont
  {Stanton}, \citenamefont {Kent}, \citenamefont {Akimov}, \citenamefont
  {Hawker}, \citenamefont {Cheng},\ and\ \citenamefont {Foxon}}]{Stanton_2001}%
  \BibitemOpen
  \bibfield  {author} {\bibinfo {author} {\bibfnamefont {N.~M.}\ \bibnamefont
  {Stanton}}, \bibinfo {author} {\bibfnamefont {A.~J.}\ \bibnamefont {Kent}},
  \bibinfo {author} {\bibfnamefont {A.~V.}\ \bibnamefont {Akimov}}, \bibinfo
  {author} {\bibfnamefont {P.}~\bibnamefont {Hawker}}, \bibinfo {author}
  {\bibfnamefont {T.~S.}\ \bibnamefont {Cheng}},\ and\ \bibinfo {author}
  {\bibfnamefont {C.~T.}\ \bibnamefont {Foxon}},\ }\bibfield  {title} {\bibinfo
  {title} {Energy relaxation by hot electrons in n-{{GaN}} epilayers},\ }\href
  {https://doi.org/10.1063/1.1334642} {\bibfield  {journal} {\bibinfo
  {journal} {J. Appl. Phys.}\ }\textbf {\bibinfo {volume} {89}},\ \bibinfo
  {pages} {973} (\bibinfo {year} {2001})}\BibitemShut {NoStop}%
\bibitem [{\citenamefont {Selbmann}\ \emph {et~al.}(1996)\citenamefont
  {Selbmann}, \citenamefont {Gulia}, \citenamefont {Rossi}, \citenamefont
  {Molinari},\ and\ \citenamefont {Lugli}}]{selbmann_1996}%
  \BibitemOpen
  \bibfield  {author} {\bibinfo {author} {\bibfnamefont {P.~E.}\ \bibnamefont
  {Selbmann}}, \bibinfo {author} {\bibfnamefont {M.}~\bibnamefont {Gulia}},
  \bibinfo {author} {\bibfnamefont {F.}~\bibnamefont {Rossi}}, \bibinfo
  {author} {\bibfnamefont {E.}~\bibnamefont {Molinari}},\ and\ \bibinfo
  {author} {\bibfnamefont {P.}~\bibnamefont {Lugli}},\ }\bibfield  {title}
  {\bibinfo {title} {Coupled free-carrier and exciton relaxation in optically
  excited semiconductors},\ }\href {https://doi.org/10.1103/physrevb.54.4660}
  {\bibfield  {journal} {\bibinfo  {journal} {Phys. Rev. B}\ }\textbf {\bibinfo
  {volume} {54}},\ \bibinfo {pages} {4660} (\bibinfo {year}
  {1996})}\BibitemShut {NoStop}%
\bibitem [{\citenamefont {Leheny}\ \emph {et~al.}(1979)\citenamefont {Leheny},
  \citenamefont {Shah}, \citenamefont {Fork}, \citenamefont {Shank},\ and\
  \citenamefont {Migus}}]{leheny_1979}%
  \BibitemOpen
  \bibfield  {author} {\bibinfo {author} {\bibfnamefont {R.~F.}\ \bibnamefont
  {Leheny}}, \bibinfo {author} {\bibfnamefont {J.}~\bibnamefont {Shah}},
  \bibinfo {author} {\bibfnamefont {R.~L.}\ \bibnamefont {Fork}}, \bibinfo
  {author} {\bibfnamefont {C.~V.}\ \bibnamefont {Shank}},\ and\ \bibinfo
  {author} {\bibfnamefont {A.}~\bibnamefont {Migus}},\ }\bibfield  {title}
  {\bibinfo {title} {Dynamics of hot carrier cooling in photo-excited
  {{GaAs}}},\ }\href {https://doi.org/10.1016/0038-1098(79)90393-4} {\bibfield
  {journal} {\bibinfo  {journal} {Solid State Commun.}\ }\textbf {\bibinfo
  {volume} {31}},\ \bibinfo {pages} {809} (\bibinfo {year} {1979})}\BibitemShut
  {NoStop}%
\bibitem [{\citenamefont {Lugli}\ \emph {et~al.}(1987)\citenamefont {Lugli},
  \citenamefont {Jacoboni}, \citenamefont {Reggiani},\ and\ \citenamefont
  {Kocevar}}]{lugli_1987}%
  \BibitemOpen
  \bibfield  {author} {\bibinfo {author} {\bibfnamefont {P.}~\bibnamefont
  {Lugli}}, \bibinfo {author} {\bibfnamefont {C.}~\bibnamefont {Jacoboni}},
  \bibinfo {author} {\bibfnamefont {L.}~\bibnamefont {Reggiani}},\ and\
  \bibinfo {author} {\bibfnamefont {P.}~\bibnamefont {Kocevar}},\ }\bibfield
  {title} {\bibinfo {title} {Monte {{Carlo}} algorithm for hot phonons in polar
  semiconductors},\ }\href {https://doi.org/10.1063/1.97925} {\bibfield
  {journal} {\bibinfo  {journal} {Appl. Phys. Lett.}\ }\textbf {\bibinfo
  {volume} {50}},\ \bibinfo {pages} {1251} (\bibinfo {year}
  {1987})}\BibitemShut {NoStop}%
\bibitem [{\citenamefont {Lobentanzer}\ \emph {et~al.}(1987)\citenamefont
  {Lobentanzer}, \citenamefont {Polland}, \citenamefont {R{\"u}hle},
  \citenamefont {Stolz},\ and\ \citenamefont {Ploog}}]{lobentanzer_1987}%
  \BibitemOpen
  \bibfield  {author} {\bibinfo {author} {\bibfnamefont {H.}~\bibnamefont
  {Lobentanzer}}, \bibinfo {author} {\bibfnamefont {H.-J.}\ \bibnamefont
  {Polland}}, \bibinfo {author} {\bibfnamefont {W.~W.}\ \bibnamefont
  {R{\"u}hle}}, \bibinfo {author} {\bibfnamefont {W.}~\bibnamefont {Stolz}},\
  and\ \bibinfo {author} {\bibfnamefont {K.}~\bibnamefont {Ploog}},\ }\bibfield
   {title} {\bibinfo {title} {Cooling of an electron-hole plasma in a
  {{Ga}}{\textsubscript{0.47}}{{In}}{\textsubscript{0.53}}{{As}}
  multiple-quantum-well structure},\ }\href
  {https://doi.org/10.1103/physrevb.36.1136} {\bibfield  {journal} {\bibinfo
  {journal} {Phys. Rev. B}\ }\textbf {\bibinfo {volume} {36}},\ \bibinfo
  {pages} {1136} (\bibinfo {year} {1987})}\BibitemShut {NoStop}%
\bibitem [{\citenamefont {Leo}\ \emph {et~al.}(1988)\citenamefont {Leo},
  \citenamefont {R{\"u}hle},\ and\ \citenamefont {Ploog}}]{leo_1988}%
  \BibitemOpen
  \bibfield  {author} {\bibinfo {author} {\bibfnamefont {K.}~\bibnamefont
  {Leo}}, \bibinfo {author} {\bibfnamefont {W.~W.}\ \bibnamefont {R{\"u}hle}},\
  and\ \bibinfo {author} {\bibfnamefont {K.}~\bibnamefont {Ploog}},\ }\bibfield
   {title} {\bibinfo {title} {Hot-carrier energy-loss rates in
  {{GaAs}}/{{Al\textsubscript{x}Ga}}{$_{1-}$}{{\textsubscript{x}As}} quantum
  wells},\ }\href {https://doi.org/10.1103/physrevb.38.1947} {\bibfield
  {journal} {\bibinfo  {journal} {Phys. Rev. B}\ }\textbf {\bibinfo {volume}
  {38}},\ \bibinfo {pages} {1947} (\bibinfo {year} {1988})}\BibitemShut
  {NoStop}%
\bibitem [{\citenamefont {Marchetti}\ and\ \citenamefont
  {P{\"o}tz}(1989)}]{marchetti_1989}%
  \BibitemOpen
  \bibfield  {author} {\bibinfo {author} {\bibfnamefont {M.~C.}\ \bibnamefont
  {Marchetti}}\ and\ \bibinfo {author} {\bibfnamefont {W.}~\bibnamefont
  {P{\"o}tz}},\ }\bibfield  {title} {\bibinfo {title} {Relaxation of
  photoexcited electron-hole plasma in quantum wells},\ }\href
  {https://doi.org/10.1103/physrevb.40.12391} {\bibfield  {journal} {\bibinfo
  {journal} {Phys. Rev. B}\ }\textbf {\bibinfo {volume} {40}},\ \bibinfo
  {pages} {12391} (\bibinfo {year} {1989})}\BibitemShut {NoStop}%
\bibitem [{\citenamefont {Shah}\ \emph {et~al.}(1970)\citenamefont {Shah},
  \citenamefont {Leite},\ and\ \citenamefont {Scott}}]{shah_1970}%
  \BibitemOpen
  \bibfield  {author} {\bibinfo {author} {\bibfnamefont {J.}~\bibnamefont
  {Shah}}, \bibinfo {author} {\bibfnamefont {R.~C.~C.}\ \bibnamefont {Leite}},\
  and\ \bibinfo {author} {\bibfnamefont {J.~F.}\ \bibnamefont {Scott}},\
  }\bibfield  {title} {\bibinfo {title} {Photoexcited hot {{LO}} phonons in
  {{GaAs}}},\ }\href {https://doi.org/10.1016/0038-1098(70)90002-5} {\bibfield
  {journal} {\bibinfo  {journal} {Solid State Commun.}\ }\textbf {\bibinfo
  {volume} {8}},\ \bibinfo {pages} {1089} (\bibinfo {year} {1970})}\BibitemShut
  {NoStop}%
\bibitem [{\citenamefont {Shah}(1986)}]{shah_1986}%
  \BibitemOpen
  \bibfield  {author} {\bibinfo {author} {\bibfnamefont {J.}~\bibnamefont
  {Shah}},\ }\bibfield  {title} {\bibinfo {title} {Hot carriers in
  quasi-2-{{D}} polar semiconductors},\ }\href
  {https://doi.org/10.1109/jqe.1986.1073164} {\bibfield  {journal} {\bibinfo
  {journal} {IEEE J. Quantum Electron.}\ }\textbf {\bibinfo {volume} {22}},\
  \bibinfo {pages} {1728} (\bibinfo {year} {1986})}\BibitemShut {NoStop}%
\bibitem [{Note2()}]{Note2}%
  \BibitemOpen
  \bibinfo {note} {Note that, in principle, surface recombination may affect
  the carrier density for low acceleration voltages, but can be safely ignored
  in the present case, since the \protect \emph {M}-plane surface of GaN is
  known to exhibit a very low surface recombination velocity \cite
  {Corfdir_2014}.}\BibitemShut {Stop}%
\bibitem [{Note3()}]{Note3}%
  \BibitemOpen
  \bibinfo {note} {The generation rate is determined according to \protect
  \citet {Wu_1978} with the parameters given in Ref.~\protect \rev@citealp
  {Jahn_2003} and the generation volume approximated by a cylinder of diameter
  $\sigma $ and a height corresponding to 75\% of the \protect \texttt {CASINO}
  energy loss distribution. The carrier density is then obtained with the
  carrier lifetimes measured by time-resolved photoluminescence spectroscopy
  [cf.\ Ref. \protect \rev@citealp {brandt_2020} (CD2) for the transients
  obtained for the barrierless (In,Ga)N/GaN QW].}\BibitemShut {Stop}%
\bibitem [{\citenamefont {Botha}\ and\ \citenamefont
  {Leitch}(1994)}]{Botha_1994}%
  \BibitemOpen
  \bibfield  {author} {\bibinfo {author} {\bibfnamefont {J.~R.}\ \bibnamefont
  {Botha}}\ and\ \bibinfo {author} {\bibfnamefont {A.~W.~R.}\ \bibnamefont
  {Leitch}},\ }\bibfield  {title} {\bibinfo {title} {Thermally activated
  carrier escape mechanisms from
  {{In{\textsubscript{x}}Ga{\textsubscript{1-x}}As}}/{GaAs} quantum wells},\
  }\href {https://doi.org/10.1103/PhysRevB.50.18147} {\bibfield  {journal}
  {\bibinfo  {journal} {Phys. Rev. B}\ }\textbf {\bibinfo {volume} {50}},\
  \bibinfo {pages} {18147} (\bibinfo {year} {1994})}\BibitemShut {NoStop}%
\bibitem [{\citenamefont {Schroeder}(1994)}]{Schroeder_1994}%
  \BibitemOpen
  \bibfield  {author} {\bibinfo {author} {\bibfnamefont {D.}~\bibnamefont
  {Schroeder}},\ }\href {https://doi.org/10.1007/978-3-7091-6644-4} {\emph
  {\bibinfo {title} {Modelling of {{Interface Carrier Transport}} for {{Device
  Simulation}}}}},\ edited by\ \bibinfo {editor} {\bibfnamefont
  {S.}~\bibnamefont {Selberherr}},\ Computational {{Microelectronics}}\
  (\bibinfo  {publisher} {{Springer Vienna}},\ \bibinfo {address} {{Vienna}},\
  \bibinfo {year} {1994})\BibitemShut {NoStop}%
\bibitem [{\citenamefont {Schneider}\ and\ \citenamefont {von
  Klitzing}(1988)}]{Schneider_1988}%
  \BibitemOpen
  \bibfield  {author} {\bibinfo {author} {\bibfnamefont {H.}~\bibnamefont
  {Schneider}}\ and\ \bibinfo {author} {\bibfnamefont {K.}~\bibnamefont {von
  Klitzing}},\ }\bibfield  {title} {\bibinfo {title} {Thermionic emission and
  gaussian transport of holes in a
  {{GaAs}}/{{Al\textsubscript{x}Ga}}{$_{1-}$}{{\textsubscript{x}As}}
  multiple-quantum-well structure},\ }\href
  {https://doi.org/10.1103/PhysRevB.38.6160} {\bibfield  {journal} {\bibinfo
  {journal} {Phys. Rev. B}\ }\textbf {\bibinfo {volume} {38}},\ \bibinfo
  {pages} {6160} (\bibinfo {year} {1988})}\BibitemShut {NoStop}%
\bibitem [{\citenamefont {Zou}\ \emph {et~al.}(1992)\citenamefont {Zou},
  \citenamefont {Rammer},\ and\ \citenamefont {Chao}}]{Zou_1992}%
  \BibitemOpen
  \bibfield  {author} {\bibinfo {author} {\bibfnamefont {N.}~\bibnamefont
  {Zou}}, \bibinfo {author} {\bibfnamefont {J.}~\bibnamefont {Rammer}},\ and\
  \bibinfo {author} {\bibfnamefont {K.~A.}\ \bibnamefont {Chao}},\ }\bibfield
  {title} {\bibinfo {title} {Tunneling escape of electrons from a
  double-barrier structure},\ }\href
  {https://doi.org/10.1103/PhysRevB.46.15912} {\bibfield  {journal} {\bibinfo
  {journal} {Phys. Rev. B}\ }\textbf {\bibinfo {volume} {46}},\ \bibinfo
  {pages} {15912} (\bibinfo {year} {1992})}\BibitemShut {NoStop}%
\bibitem [{\citenamefont {Nash}\ \emph {et~al.}(1987)\citenamefont {Nash},
  \citenamefont {Skolnick},\ and\ \citenamefont {Bass}}]{nash_1987}%
  \BibitemOpen
  \bibfield  {author} {\bibinfo {author} {\bibfnamefont {K.~J.}\ \bibnamefont
  {Nash}}, \bibinfo {author} {\bibfnamefont {M.~S.}\ \bibnamefont {Skolnick}},\
  and\ \bibinfo {author} {\bibfnamefont {S.~J.}\ \bibnamefont {Bass}},\
  }\bibfield  {title} {\bibinfo {title} {Electron-phonon interactions in indium
  gallium arsenide},\ }\href {https://doi.org/10.1088/0268-1242/2/6/002}
  {\bibfield  {journal} {\bibinfo  {journal} {Semicond. Sci. Technol.}\
  }\textbf {\bibinfo {volume} {2}},\ \bibinfo {pages} {329} (\bibinfo {year}
  {1987})}\BibitemShut {NoStop}%
\bibitem [{\citenamefont {Barker}\ and\ \citenamefont
  {Ilegems}(1973)}]{barker_1973}%
  \BibitemOpen
  \bibfield  {author} {\bibinfo {author} {\bibfnamefont {A.~S.}\ \bibnamefont
  {Barker}}\ and\ \bibinfo {author} {\bibfnamefont {M.}~\bibnamefont
  {Ilegems}},\ }\bibfield  {title} {\bibinfo {title} {Infrared lattice
  vibrations and free-electron dispersion in {{GaN}}},\ }\href
  {https://doi.org/10.1103/physrevb.7.743} {\bibfield  {journal} {\bibinfo
  {journal} {Phys. Rev. B}\ }\textbf {\bibinfo {volume} {7}},\ \bibinfo {pages}
  {743} (\bibinfo {year} {1973})}\BibitemShut {NoStop}%
\bibitem [{\citenamefont {Prabhu}\ \emph {et~al.}(1995)\citenamefont {Prabhu},
  \citenamefont {Vengurlekar}, \citenamefont {Roy},\ and\ \citenamefont
  {Shah}}]{prabhu_1995}%
  \BibitemOpen
  \bibfield  {author} {\bibinfo {author} {\bibfnamefont {S.~S.}\ \bibnamefont
  {Prabhu}}, \bibinfo {author} {\bibfnamefont {A.~S.}\ \bibnamefont
  {Vengurlekar}}, \bibinfo {author} {\bibfnamefont {S.~K.}\ \bibnamefont
  {Roy}},\ and\ \bibinfo {author} {\bibfnamefont {J.}~\bibnamefont {Shah}},\
  }\bibfield  {title} {\bibinfo {title} {Nonequilibrium dynamics of hot
  carriers and hot phonons in {{CdSe}} and {{GaAs}}},\ }\href
  {https://doi.org/10.1103/physrevb.51.14233} {\bibfield  {journal} {\bibinfo
  {journal} {Phys. Rev. B}\ }\textbf {\bibinfo {volume} {51}},\ \bibinfo
  {pages} {14233} (\bibinfo {year} {1995})}\BibitemShut {NoStop}%
\bibitem [{\citenamefont {Corfdir}\ \emph {et~al.}(2014)\citenamefont
  {Corfdir}, \citenamefont {Hauswald}, \citenamefont {Zettler}, \citenamefont
  {Flissikowski}, \citenamefont {L{\"a}hnemann}, \citenamefont
  {{Fern{\'a}ndez-Garrido}}, \citenamefont {Geelhaar}, \citenamefont {Grahn},\
  and\ \citenamefont {Brandt}}]{Corfdir_2014}%
  \BibitemOpen
  \bibfield  {author} {\bibinfo {author} {\bibfnamefont {P.}~\bibnamefont
  {Corfdir}}, \bibinfo {author} {\bibfnamefont {C.}~\bibnamefont {Hauswald}},
  \bibinfo {author} {\bibfnamefont {J.~K.}\ \bibnamefont {Zettler}}, \bibinfo
  {author} {\bibfnamefont {T.}~\bibnamefont {Flissikowski}}, \bibinfo {author}
  {\bibfnamefont {J.}~\bibnamefont {L{\"a}hnemann}}, \bibinfo {author}
  {\bibfnamefont {S.}~\bibnamefont {{Fern{\'a}ndez-Garrido}}}, \bibinfo
  {author} {\bibfnamefont {L.}~\bibnamefont {Geelhaar}}, \bibinfo {author}
  {\bibfnamefont {H.~T.}\ \bibnamefont {Grahn}},\ and\ \bibinfo {author}
  {\bibfnamefont {O.}~\bibnamefont {Brandt}},\ }\bibfield  {title} {\bibinfo
  {title} {Stacking faults as quantum wells in nanowires: Density of states,
  oscillator strength, and radiative efficiency},\ }\href
  {https://doi.org/10.1103/PhysRevB.90.195309} {\bibfield  {journal} {\bibinfo
  {journal} {Phys. Rev. B}\ }\textbf {\bibinfo {volume} {90}},\ \bibinfo
  {pages} {195309} (\bibinfo {year} {2014})}\BibitemShut {NoStop}%
\bibitem [{\citenamefont {Wu}\ and\ \citenamefont {Wittry}(1978)}]{Wu_1978}%
  \BibitemOpen
  \bibfield  {author} {\bibinfo {author} {\bibfnamefont {C.~J.}\ \bibnamefont
  {Wu}}\ and\ \bibinfo {author} {\bibfnamefont {D.~B.}\ \bibnamefont
  {Wittry}},\ }\bibfield  {title} {\bibinfo {title} {Investigation of
  minority-carrier diffusion lengths by electron bombardment of {{Schottky}}
  barriers},\ }\href {https://doi.org/10.1063/1.325163} {\bibfield  {journal}
  {\bibinfo  {journal} {J. Appl. Phys.}\ }\textbf {\bibinfo {volume} {49}},\
  \bibinfo {pages} {2827} (\bibinfo {year} {1978})}\BibitemShut {NoStop}%
\bibitem [{\citenamefont {Jahn}\ \emph {et~al.}(2003)\citenamefont {Jahn},
  \citenamefont {Dhar}, \citenamefont {Brandt}, \citenamefont {Grahn},
  \citenamefont {Ploog},\ and\ \citenamefont {Watson}}]{Jahn_2003}%
  \BibitemOpen
  \bibfield  {author} {\bibinfo {author} {\bibfnamefont {U.}~\bibnamefont
  {Jahn}}, \bibinfo {author} {\bibfnamefont {S.}~\bibnamefont {Dhar}}, \bibinfo
  {author} {\bibfnamefont {O.}~\bibnamefont {Brandt}}, \bibinfo {author}
  {\bibfnamefont {H.~T.}\ \bibnamefont {Grahn}}, \bibinfo {author}
  {\bibfnamefont {K.~H.}\ \bibnamefont {Ploog}},\ and\ \bibinfo {author}
  {\bibfnamefont {I.~M.}\ \bibnamefont {Watson}},\ }\bibfield  {title}
  {\bibinfo {title} {Exciton localization and quantum
  efficiency\textemdash{{A}} comparative cathodoluminescence study of
  ({{In}},{{Ga}}){{N}}/{{GaN}} and {{GaN}}/({{Al}},{{Ga}}){{N}} quantum
  wells},\ }\href {https://doi.org/10.1063/1.1529993} {\bibfield  {journal}
  {\bibinfo  {journal} {J. Appl. Phys.}\ }\textbf {\bibinfo {volume} {93}},\
  \bibinfo {pages} {1048} (\bibinfo {year} {2003})}\BibitemShut {NoStop}%
\end{thebibliography}%

\end{document}